\documentclass[12pt]{article}

\usepackage[utf8]{inputenc}   % UTF‑8 support
\usepackage[T1]{fontenc}

\RequirePackage[tt=false, type1=true]{libertine} 
\RequirePackage[varqu]{zi4}

\usepackage{geometry}         % Easy margins
\usepackage{microtype}        % Better kerning & spacing

\usepackage{tikz}
\usepackage{standalone}

\usepackage{graphicx}%
\usepackage{multirow}%
\usepackage{amsmath,amsfonts}%
\usepackage{amssymb}
\usepackage{xcolor}%
\usepackage{textcomp}%
\usepackage{manyfoot}%
\usepackage{booktabs}%
\usepackage{listings}
\usepackage{tabularx} % in the preamble
\usepackage{caption}

\usepackage[ruled, vlined,linesnumbered]{algorithm2e}
\usepackage{xspace}
\usepackage[inline]{enumitem} 
\usepackage{multirow}
\usepackage{pifont}
\usepackage{balance}
\usepackage{comment}
\usepackage{graphics}
\usepackage[normalem]{ulem}
\usepackage{float}

\usepackage{commands}
\usepackage{amsthm}

\RequirePackage[libertine]{newtxmath}

\usepackage[bookmarks]{hyperref}
\hypersetup{
	colorlinks=true,
	linkcolor=myblue,
	citecolor=colorcite,
	urlcolor=colorcite
}

\usepackage[nameinlink,capitalize]{cleveref}
\usepackage[sort&compress]{natbib}

\newtheorem{definition}{Definition}
\newtheorem{problem}{Problem}
\newtheorem{lemma}{Lemma}
\newtheorem{theorem}{Theorem}
\newtheorem{corollary}{Corollary}
\DeclareMathOperator{\expectation}{\mathbb{E}}
\DeclareMathOperator{\Prob}{\mathbb{P}}
\DeclareMathOperator{\Var}{Var}
\DeclareMathOperator{\Cov}{Cov}
\DeclareMathOperator{\Unif}{\mathfrak{U}}

\title{Efficient Approximate Temporal Triangle Counting\\ in Streaming with Predictions}

\author{%
	\hspace{-1cm}
	\begin{tabular}{lc}
		\begin{tabular}{c}
			Giorgio Venturin\thanks{Equal contribution.} \\ \url{giorgio.venturin@phd.unipd.it} \\ University of Padova 
		\end{tabular} & 
		\begin{tabular}{c}
			Ilie Sarpe$^*$ \\ \url{ilsarpe@kth.se} \\ KTH Royal Institute of Technology
		\end{tabular}\\
	\end{tabular}\\[4ex]
	\centering
	\begin{tabular}{c}
		\begin{tabular}{c}
			Fabio Vandin \\ \url{fabio.vandin@unipd.it}  \\ University of Padova
		\end{tabular}\\
	\end{tabular}
}

\date{}

\begin{document}

\maketitle

\begin{abstract}
  Triangle counting is a fundamental and widely studied problem on static graphs, and recently on \emph{temporal graphs}, where edges carry information on the timings of the associated events. \emph{Streaming} processing and resource \emph{efficiency} are crucial requirements for counting triangles in modern massive temporal graphs, with millions of nodes and up to billions of temporal edges.
  
  However, current exact and approximate algorithms are unable to handle large-scale temporal graphs. 
  To fill such a gap, we introduce \algName, a scalable and efficient algorithm to approximate temporal triangle counts from a stream of temporal edges. \algName combines \emph{predictions} to the number of triangles a temporal edge is involved in, with a simple sampling strategy, leading to scalability, efficiency, and accurate approximation of all eight temporal triangle types simultaneously. We analytically prove that, by using a sublinear amount of memory, \algName obtains unbiased and very accurate estimates. In fact, even noisy predictions can significantly reduce the variance of \algName's estimates. 
  
  Our extensive experiments on massive temporal graphs with up to billions of edges demonstrate that \algName outputs high-quality estimates and is more efficient than state-of-the-art methods.
\end{abstract}

\newpage
\section{Introduction}\label{intro}

\emph{Temporal graphs} model complex time-evolving systems~\citep{Holme2012TNSurvey}, including social networks~\citep{Tang2009Socials}, e-commerce platforms~\citep{Gao2023Survey}, databases~\citep{Debrouvier2021TemDat}, and biological systems~\citep{Hulovatyy2015Temporal}, by associating each event in the system with its \emph{timing} of occurrence. Temporal graph analysis provides a \emph{deep understanding} of the underlying complex systems and their properties~\citep{Holme2023Book} through various problems such as temporal reachability~\citep{Wu2014TemporalPaths}, temporal communities~\citep{Lin2024TemporalCommunity}, databases \citep{Hu2022TemporalJoins,Hou2024TempDatabase}, core decomposition~\citep{Qin2022BurstyCore}, and more~\citep{Gionis2024Tutorial}.

\emph{Temporal motifs} and temporal triangles~\citep{Liu2021MotifsDefs,paranjape2017motifs} are fundamental patterns defined by $i$) a subgraph representing a given \emph{topological property}, $ii$) an ordering over the edges, capturing the timing of occurrence of the subgraph edges, and $iii$) a temporal proximity constraint assuring that all events occur close in time.
Temporal motifs, especially temporal triangles, and their \emph{counts} (i.e., the number of occurrences of a temporal motif in a temporal graph), are crucial for the analysis of temporal graphs. Some applications include graph classification~\citep{Tu2019gl2vec}, anomaly detection~\citep{Belth2020Persistence}, fraud detection~\citep{Wu2022Mixing,Liu2024Fishing}, travel pattern analysis~\citep{Lei2020Travel}, dense subgraph identification~\citep{Sarpe2024TMDS}, synthetic network generation~\citep{Porter2022SBM}, and more~\citep{Gionis2024Tutorial}.
This is analogous to the wide use of motifs and triangles to analyze static graphs~\citep{Seshadhri2019Tutorial}---but the temporal dimension often poses additional challenges compared to static graphs. For instance, identifying a single star-shaped temporal motif is \(\mathbf{NP}\)-hard~\citep{liu2019sampling}, unlike static graphs where counting all star-shaped subgraphs can be done in polynomial time~\citep{Seshadhri2019Tutorial}.

For the problem of \emph{temporal triangle counting}, both exact and approximate algorithms have been developed~\citep{Mackey2018EdgeDriven,pashanasangi2021faster,sarpe2021oden}. However, exact approaches do not scale to modern-sized temporal graphs~\citep{paranjape2017motifs,gao2022scalable,Mackey2018EdgeDriven,li2024motto,pashanasangi2021faster}. Furthermore, approximate methods are based on sampling techniques requiring very large and impractical resources (time and memory) to obtain provably accurate estimates, due to their worst-case assumptions on the input~\citep{wang2022efficient,sarpe2021presto,liu2019sampling}. Overall, both exact and approximate methods still require substantial resources to process large temporal graphs, and designing scalable algorithms for temporal triangle counting remains a challenging open problem.

\para{Main contributions.} We introduce \algName, a new algorithm for counting temporal triangles in large temporal graphs. \algName solves the counting problem \emph{approximately} within a challenging and practical \emph{streaming} scenario, where edges are processed in a single pass over the stream~\citep{mcgregor2014graph}. We consider the streaming approach since it addresses the challenges of analyzing massive temporal graphs, characterized by a high volume of interactions recorded over time and large memory required to store the data~\citep{Holme2012TNSurvey,Holme2023Book}. \algName, at its core, uses a randomized sampling approach coupled with the information provided by a suitable \emph{predictor} to identify and retain the most important edges over the stream. We prove that our design yields accurate estimates with sublinear memory and rigorous concentration guarantees, i.e., the output is a relative $\varepsilon$-approximation for small $\varepsilon>0$. Our extensive experimental evaluation shows that \algName saves up to $19\times$ memory and $200\times$ time compared to existing state-of-the-art methods while delivering highly accurate estimates. Our key contributions are as follows:

\begin{enumerate}
 	\item We propose \algName, a randomized, single-pass streaming algorithm that leverages a simple sampling approach coupled with a predictor to accurately estimate all temporal triangle counts simultaneously with low running time and memory usage.
        \item We rigorously analyze \algName, demonstrating that:
            	$i$) it produces unbiased estimates independent of the prediction quality,
            	$ii$) the predictor significantly decreases the variance of the estimates, and
            	$iii$) estimates remain robust and of high quality even under noisy predictions.
        \item  We design a practical and efficient predictor that, despite being domain-agnostic, can be used within \algName to enable high accuracy and small memory usage.
        \item We perform an extensive experimental evaluation on large temporal graphs to evaluate \algName and show that: $i$) it outperforms state-of-the-art (\sota) methods for temporal triangle counting, achieving highly accurate results while reducing resource usage by orders of magnitude; $ii$) \algName is the only method capable of obtaining accurate approximations on a three-billion-edge temporal graph; $iii$) \algName works in an online setting, using a predictor learned from historical data.
\end{enumerate}

\begin{figure}[t]
	\centering
	\includegraphics[width=\columnwidth]{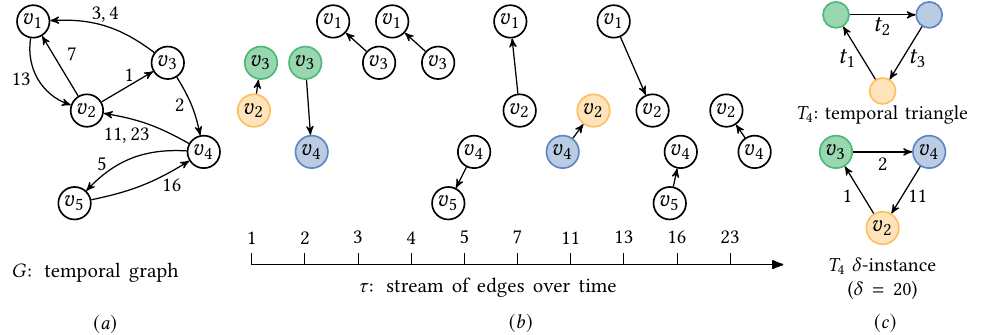}
	\includegraphics[width=\columnwidth]{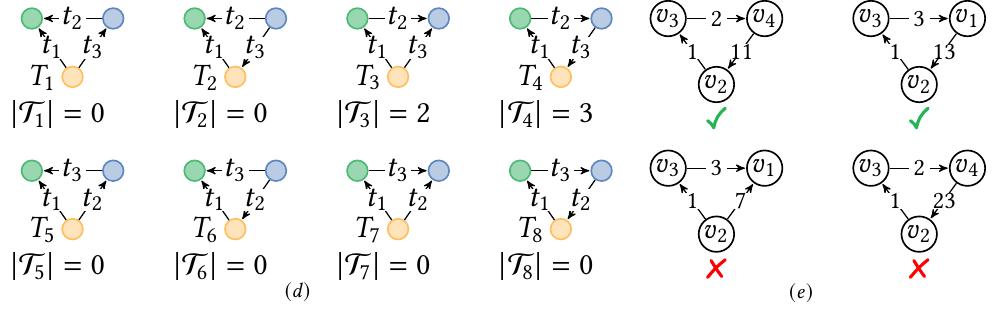}
	\caption{$(a)$: temporal graph $G=(V,E)$, with $V=\{v_1,\dots,v_5\}$ and temporal edges $E=\{(v_2,v_3,1),\dots\}$: each temporal edge is a triplet $(u,v,t)$ with $t$ being the time of occurrence. 
		$(b)$: stream over time of the temporal edges of $G$. 
		$(c)$: the sequence $S = \langle (v_2, v_3), (v_3, v_4), (v_4, v_2) \rangle$ is a $\delta$-instance of temporal triangle $T_4$ ($\delta=20$). 
		$(d)$: all distinct temporal triangles and the number of their $\delta$-instances in $G$ ($\delta=20$). Edge labels $t_i, i=1,2,3$ (with $t_1 < t_2 < t_3$) denote the ordering of the edge in the sequence $\sigma$ (see \Cref{def:tempTri}). 
		$(e)$: some sequences of edges from $G$, $\delta$-instances of triangle $T_4$ for $\delta = 20$ are marked with \vmark, while sequences with \xmark do not respect \Cref{def:deltaInst}.}
	\label{fig:triangles}
\end{figure}

\section{Preliminaries} 
\label{prelims}

We start by introducing the key definitions and concepts used throughout our work. 

 \begin{definition}~\label{def:tempGraph}
A temporal graph is a pair $G = (V,E) $ where $V = \{v_1,\dots, v_n\}$ is a set of $n$ vertices and $E=\{(u_1,v_1,t_1),\dots, (u_m, v_m, t_m) : u_i, v_i \in V, u_i\neq v_i \text{ and } t_i \in \mathbb{R^{+}} \}$ is a set of $m$ directed temporal edges. Each temporal edge $e=(u,v,t)\in E$ has a \emph{timestamp} $t\in \mathbb{R}^+$ denoting the timing of occurrence of the (static) interaction $(u,v)$. 
\end{definition}

\Cref{fig:triangles}(a) shows an example of a temporal graph $G$ with $n=5$ vertices and $m=10$ temporal edges. We use the term \emph{static edge} to denote an edge $(u,v)\in V\times V$ with no associated timestamp. Similarly, we denote a graph formed by static edges as a \emph{static} graph.\footnote{The static graph of $G$ is obtained by considering all its edges as static.} We are interested in counting \emph{temporal triangles}~\citep{paranjape2017motifs,liu2019sampling}, fundamental patterns for analyzing temporal graphs and defined as follows.

\begin{definition}[\citep{paranjape2017motifs,liu2019sampling}]\label{def:tempTri}
A \emph{temporal triangle} is a pair $T=((V_{T}, E_{T}), \sigma)$ where $(V_T, E_T)$ is a static graph with $|V_T|=3$ vertices and  $|E_T| = 3$ edges, and $\sigma$  is an ordering of the edges in $E_{T}$.
\end{definition}

A temporal triangle $T=((V_T,E_T),\sigma)$ captures \emph{both} structural and temporal properties, since: 
	($i$) the directed triangle $(V_T, E_T)$ represents a triadic interaction, and
	($ii$) the ordering $\sigma$ captures a \emph{temporal ordering}, i.e.,  how the sequence of edges of $T$ occurs in time. 
Note that there are eight \emph{distinct} temporal triangles: we denote each temporal triangle with $T_i, i \in [8]$\footnote{We use $i \in [a], a\in \mathbb{N}$ to denote $i \in \{1, \dots, a\}$.} (see \Cref{fig:triangles}(d)), while we use $T$ to denote an arbitrary temporal triangle. We now formalize the notion of occurrence of a temporal triangle $T$ in a temporal graph.

\begin{definition}\label{def:deltaInst}
Let $T=((V_T,E_T),\sigma)$ be a temporal triangle where $\langle(x_{1}, y_{1}), (x_{2}, $ $y_{2}),(x_{3}, y_{3})\rangle$ is the sequence of edges of $E_T$ ordered according to $\sigma$. 
Given a temporal graph $G=(V,E)$ and a time duration $\delta\in \mathbb{R}^+$ we say that a sequence of temporally-ordered edges $S = \langle (u_{1}, v_{1}, t_{1}), (u_{2}, v_{2}, t_{2}),$ $ (u_{3}, v_{3}, t_{3}) \rangle$ from $E$ is a \emph{$\delta$-instance} of the temporal triangle $T$ if:
		1) there exists a bijection $f$ on the vertices such that $f(u_{i}) = x_{i}, f(v_{i}) = y_{i}$ for $i \in \{1,2,3\}$;
		2) the time-duration of the sequence $S$ is at most $\delta$, that is $t_{3} - t_{1} \le \delta$.
\end{definition}

Given a time-duration $\delta\in \mathbb{R}^+$, a $\delta$-instance $S$ represents an \emph{occurrence} of the temporal triangle $T$: \emph{1)} the sequence $S$ of three edges from $E$ maps on $(V_T, E_T)$ following the temporal order given by $\sigma$; and \emph{2)} all edges of $S$ co-occur sufficiently close within $\delta$-time, accounting for temporal proximity. See \Cref{fig:triangles}(c) and \Cref{fig:triangles}(e)  for detailed examples. Given a temporal graph $G$, a temporal triangle $T_i, i\in[8]$, and a time-duration $\delta$, we let $\triType_{i}  = \{\tri \in E^3 : \tri$ is a $\delta$-instance of $T_i$ in $G \}$ be the \emph{set} of $\delta$-instances of $T_i$ in $G$. For $i\in[8]$, we define the \emph{count} of triangle $T_i$ as $|\triType_{i}|$. Note that for a temporal graph with $m$ temporal edges, the count $|\triType_{i}|$ of triangle $T_i$ can be as large as $\bigT(m^3)$, and, in contrast to static graphs, $m$ may not be polynomial in $n$ (due to the edges timestamps).

\spara{Streaming model.} We focus on the restrictive \emph{streaming} computational model, consisting of the following constraints when processing the temporal graph $G$: 

\begin{enumerate}
\item the temporal graph $G$ is made available as a stream $\tau$ of temporal edges;
\item temporal edges in the stream $\tau$ are \emph{temporally ordered}, such that if $e_1 = (u_1, v_1, t_1)$ precedes $e_2 = (u_2, v_2, t_2)$ in $\tau$, then $t_1 < t_2$; and 
\item each temporal edge can be processed only \emph{once}, in a single 1-pass over the stream $\tau$. 
\end{enumerate}
 
See \Cref{fig:triangles}(b) for an example of a stream of a temporal graph. Note that the streaming model above is significantly more challenging than most computational models adopted in prior works. In particular, existing methods for counting temporal
subgraphs, such as motifs or triangles, allow for \emph{multiple passes} over $\tau$~\citep{wang2022efficient}, or \emph{complete random access} to the graph~\citep{pashanasangi2021faster,paranjape2017motifs,gao2022scalable,li2024motto}.
Our choice of the streaming model is motivated by the fact that modern temporal graphs have massive sizes, e.g., they are collected from high-throughput systems such as IP-networks or social networks~\citep{Holme2012TNSurvey}. Hence, the streaming access model is a challenging and restrictive computational model for processing temporal graphs, but of high practical utility.

\spara{Computational problem.} Given a temporal graph $G$, our goal is to compute the counts $|\triType_{i}|$ for all triangles $T_i$, $i \in [8]$, simultaneously. Given that the exact computation of all the temporal triangle counts is extremely challenging, inefficient, and resource demanding~\citep{pashanasangi2021faster,Mackey2018EdgeDriven,paranjape2017motifs,gao2022scalable}, we focus on computing high-quality estimates of all counts $|\triType_{i}|, i\in[8]$, as formalized by the following problem.

\begin{problem}[Temporal triangle estimation problem]\label{prob:ourProb} Given a 1-pass stream $\tau$ of a temporal graph $G$, a time-duration $\delta\in \mathbb{R}^+$,  an approximation error $\varepsilon > 0$, and a small constant $\eta$, output estimates $c_i, i \in [8]$ such that $\mathbb{P}[|c_i - |\triType_i||\ge \varepsilon |\triType_i|] \le  \eta, \forall \ \estimate_i, i \in [8]$.
\end{problem}

\Cref{prob:ourProb} requires the \emph{simultaneous} computation of estimates $c_i$ for triangle counts $|\triType_i|, i \in [8]$, with guaranteed accuracy (i.e., relative $\varepsilon$-approximation) and bounded error probability $\eta$, in the challenging setting of a 1-pass stream. In addition, we require that an algorithm for \Cref{prob:ourProb} must use \emph{limited total memory}, since temporal triangle counting is extremely memory-demanding~\citep{Mackey2018EdgeDriven,sarpe2021presto,wang2022efficient}. Restricting the total memory to be sublinear is very common in streaming settings~\citep{wang2017approximately,muthukrishnan2005data}. For \Cref{prob:ourProb}, we focus on requiring a total memory sublinear in $m_{\delta}$, i.e., the maximum number of temporal edges of the stream $\tau$ that occur in any time window of length $\delta$.

\section{\algName algorithm}

We now introduce our algorithm \algName ($\mathtt{S}$treaming-based temporal $\mathtt{T}$riangles counts $\mathtt{E}$stimation with $\mathtt{P}$redictions). We first provide an overview of \algName (\Cref{subsec:overviewAlg}), then describe our algorithm in detail (\Cref{sec:alg}). We then discuss the theoretical guarantees of~\algName and relate the accuracy of its estimates with the quality of the predictions (\Cref{sec:oracle}), and conclude by describing a simple and practical predictor for~\algName (\Cref{sec:mdp}). All missing proofs and subroutines are in \Cref{appsec:missingProofs} and \Cref{appendix::subroutines}.

\subsection{Overview}\label{subsec:overviewAlg}

We start by introducing an overview of the techniques and the design choices behind our algorithm~\algName. We design~\algName to achieve subliner memory guarantees (see \Cref{sec:oracle}). To do so, \algName builds on ideas from state-of-the-art streaming algorithms for sublinear counting of \emph{static} triangles and subgraphs~\citep{Seshadhri2019Tutorial}. That is:

\emph{1)} \algName stores, probabilistically, a \emph{small sample} of edges from the stream $\tau$; 

\emph{2)} \algName computes unbiased estimates $\estimate_i$ \emph{simultaneously} for each count $|\triType_{i}|, i\in [8]$ based on the \emph{random} sample obtained above.

The above approach allows~\algName to use sublinear memory and to obtain \emph{unbiased} estimates $\estimate_i$, i.e., the expectation of $\estimate_i$  is equal to the count $|\triType_{i}|$.
However, the actual estimates $\estimate_i$ may be very far from $|\triType_{i}|$,  
especially when the random sample retained by \algName is not representative. 
To enable~\algName to compute estimates $c_i$ close to their expectations $|\triType_{i}|$, 
we build on ideas from the Algorithms-with-Predictions literature for \emph{static graphs}~\citep{chen2022triangle, boldrin2024fast}: 

\begin{itemize}
\item we empower \algName with a predictor $\oracle{\cdot}$ that enables the identification of important edges on the stream $\tau$---yielding representative samples retained by \algName and hence very accurate estimates $c_i$;

\item we design a predictor for the \emph{simultaneous estimation} of all temporal triangle counts, relating~\algName's accuracy with the quality of predictions of $\oracle{\cdot}$.
\end{itemize}

We prove that perfect predictions (in \Cref{theorem:varianceNoNoise}), and noisy predictions (in \Cref{theo:noisyGuarantees}) can yield much more accurate estimates $c_i$ compared to a sampling algorithm not using predictions. Moreover, we also design a \emph{practical} predictor $\oracle{\cdot}$ for~\algName.

\begin{algorithm}[t]
	\KwIn{Stream $\tau$ of temporal edges, time-duration $\delta$, predictor \oracle{\cdot}, sampling probability $p\in(0,1]$.}
	\KwOut{Estimates $\estimate_i$  of $|\triType_i|$ for $ i \in [8]$.}
	$H \leftarrow \emptyset$; $S_L \leftarrow \emptyset$\; \label{code:init_sets}
	$\estimate_{i,0} \leftarrow 0$; $\estimate_{i,1} \leftarrow 0$; $\estimate_{i,2} \leftarrow 0$  \text{for} $i \in [8]$\; \label{code:init_counters}
	\ForEach{$e = (u,v,t) \in \tau$}{ \label{code:stream_for_loop}
		$H \gets \mathtt{CleanUp}(H, t-\delta); S_L \gets\mathtt{CleanUp}(S_L, t-\delta)$\; \label{code:time_pruning}
		$\wedgeSet_{H,H}, \wedgeSet_{H,S_L}, \wedgeSet_{S_L,S_L} \gets \mathtt{CollectWedges}(H, S_L, e)$\; \label{code:collect_wedges}
		$\estimate_{i,0} \leftarrow \updateRoutine(\estimate_{i,0}, \wedgeSet_{S_L,S_L} , e)$ \text{for} $i \in [8]$\; \label{code:update_estimate_1}
		$\estimate_{i,1} \leftarrow \updateRoutine(\estimate_{i,1}, \wedgeSet_{H,S_L} , e)$ \text{for} $i \in [8]$\; \label{code:update_estimate_2}
		$\estimate_{i,2} \leftarrow \updateRoutine(\estimate_{i,2}, \wedgeSet_{H,H} , e)$ \text{for} $i \in [8]$\; \label{code:update_estimate_3}
		
		\lIf{$\oracle{e} = 1$}{ \label{code:heavy_classification}
			$H \leftarrow H \cup \{e\}$}
		\lElse{\textbf{if} $\text{\emph{\FlipBiasedCoin}}(p)=true$ \textbf{then}
			$S_L \leftarrow S_L \cup \{e\}$} \label{code:light_edge_sampling}
	}
	\KwRet $\estimate_i = \frac{\estimate_{i,0}}{p^2} + \frac{\estimate_{i,1} }{p}+ \estimate_{i,2}$ \textup{for} $i \in [8]$; \label{code:output}
	\caption{\algName}\label{alg:MainAlgorithm}
\end{algorithm}

\subsection{Algorithm description}
\label{sec:alg}

We now present our algorithm \algName. \algName leverages a randomized approach by sampling edges over the stream with a fixed probability $p\in (0,1]$, similarly to state-of-the-art methods for estimating temporal subgraph counts~\citep{liu2019sampling,sarpe2021presto,wang2022efficient}, but in addition it employs a \emph{predictor} $\oracle{\cdot}$, that classifies edges on the stream $\tau$ as either \emph{heavy} or \emph{light}. Heavy edges are those \emph{predicted} to be involved in many temporal triangles, and thus important to retain to collect a representative sample. Carefully leveraging the edge classification provided by $\oracle{\cdot}$ is crucial to obtain strong guarantees on small memory usage and high estimation accuracy. 

More in details, \algName works as follows. First, it initializes two sets, $H$ and $S_L$, for storing heavy and (sampled) light edges, respectively (\Cref{code:init_sets}). It then initializes counters $\estimate_{i,0}, \estimate_{i,1}, \estimate_{i,2}$  for each triangle type $T_i, i \in [8]$  (\Cref{code:init_counters}). All counters are used to output the unbiased estimates $\estimate_i$ for $|\triType_i|$. \algName then processes the stream $\stream$ (\Cref{code:stream_for_loop}), and for each edge $e=(u,v,t)$:
\begin{enumerate}
	\item it removes edges from $H$ and $S_L$ with timestamps smaller than $t - \delta$ using the \texttt{CleanUp} procedure (line~\ref{code:time_pruning}). The \texttt{CleanUp} procedure simply updates the sets $H$ and $S_L$ by retaining only edges $e' = (u',v',t')$ with $t'$ within $\delta$ time from the timestamp of the current edge $e$, that is, $t - t' \leq \delta$;
	\item  it collects all \emph{wedges} in the sample $H \cup S_L$,\footnote{A wedge is a pair of edges sharing a vertex.} since $e$ can form $\delta$-instances with all such wedges;
	\item all the collected wedges are partitioned into three subsets via the \wedgeRoutine function (\Cref{code:collect_wedges}): $\wedgeSet_{H,H}$ (wedges with both edges in $H$), $\wedgeSet_{S_L,S_L}$ (both edges in $S_L$), and $\wedgeSet_{H, S_L}$ (one edge in $H$, the other in $S_L$);
	\item the counters $\estimate_{i,0}, \estimate_{i,1}, \estimate_{i,2}$ are updated using the \updateRoutine procedure (\Cref{code:update_estimate_1}--\Cref{code:update_estimate_3}), tracking occurrences of triangles $T_i$ with $0$, $1$, or $2$ heavy edges. 
\end{enumerate}

\algName then calls the predictor $\oracle{e}$ to determine whether $e$ is heavy or light: if $\oracle{e}=1$, $e$ is added to $H$ (\Cref{code:heavy_classification}); otherwise, $e$ is added to $S_L$ with probability $p$ using the \FlipBiasedCoin function (\Cref{code:light_edge_sampling}). Finally, \algName outputs estimates $\estimate_i$ for $i \in [8]$ by combining and weighting the counters $c_{i,j}, j=0,1,2$ (\Cref{code:output}).

\subsection{Analysis}\label{subsec:oraclesAndVariance}
\label{sec:oracle}

\para{Time complexity.} First, we briefly consider the \emph{expected} time complexity of \algName. Given the input to \Cref{alg:MainAlgorithm}, assuming the predictions from $\oracle{\cdot}$ require constant time,\footnote{This is achieved in practice for the predictor of \Cref{sec:mdp} using hashing.} the expected time complexity of \algName is $\bigO(m (pm_{\delta} + |H|)^2)$, 
where: $m_\delta$ is the maximum number of edges over $\tau$ that occur in any time-duration of length $\delta$, and $|H|$ is the maximum number of heavy edges in $H$ during the execution of the algorithm (see \Cref{appendix:time_complexity} for a detailed analysis). Note that when we select $|H| = o(m_\delta)$ and $p \ll 1$  (as done in our experiments), \algName becomes much more efficient than previous approaches (see \Cref{sec:related_works}). In practice, such complexity enables~\algName to scale to large datasets, where previous approaches become impractical (see our experiments in \Cref{sec:experiments}). 

\smallskip
\para{Unbiasedness.} We prove that \algName computes \emph{unbiased estimates} $\estimate_i$ of the counts $|\triType_i|$, for $i \in [8]$,  \emph{independently} of the quality of the predictions of $\oracle{\cdot}$.

\begin{lemma}\label{unbiasness_theorem}
Given a stream $\tau$ of a temporal graph $G=(V,E)$, a time-duration $\delta$, and a heaviness predictor \oracle{\cdot}, each estimate $\estimate_i, i \in [8]$ reported by \algName is an unbiased estimate of the count $|\triType_i|$, that is $\mathbb{E}[\estimate_i] = |\triType_i|$.
\end{lemma}
The proof is found in~\Cref{appsec:missingProofs}.

\spara{Embedding predictions.} We now analyze the impact of the predictor $\oracle{\cdot}$ for our algorithm~\algName. We first propose a practical model for a predictor, formalizing
a \emph{ranking} predictor. Our model is inspired by the empirical evidence that most machine learning models are highly optimized for ranking metrics such as Kendall's tau, Spearman's correlation, and Recall@$K$~\citep{Zhang2024TATKC,Heeg2023TemporalCentrality,Zhou2020SurveyGNN,Yu2021RankingUsers}---hence, our model is designed to account for the output of such methods. More in detail, a ranking predictor ranks edges $e\in\tau$ according to their importance for the counts $|\triType_{i}|$. 

We show analytically that a \emph{perfect} ranking predictor yields both very \emph{accurate} estimates and \emph{sublinear space complexity} for \algName. In addition, using a predictor significantly reduces the variance of the estimates computed by~\algName. Finally, we relate~\algName's memory usage and accuracy with a  \emph{noisy} predictor, showing that when the noise is not too large, good performances can still be achieved by~\algName.

\emph{Perfect predictor.} Let $\edgeWeight{e}{i} = |\{\tri : e \in \tri, \tri \in \triType_i \}|$ be the number of triangles in $\triType_i, i\in[8]$ containing edge $e\in E$, and $\totEdgeWeight{e} = \sum_{i\in[8]} \edgeWeight{e}{i}$ be the total edge weight of $e$, i.e., the total count of triangles in which edge $e$ is contained. Let $\triangleVec \doteq \langle e^\totEdgeW_1,\dots, e^\totEdgeW_{m} \rangle$ be the edges in $E$ ordered by \emph{non-increasing} values according to their weights $\totEdgeWeight{e}$ with ties broken arbitrarily. Given two distinct edges $e,e'\in E$, we use $e\prec e'$ to denote that $e$ comes before $e'$ in the ordering $\triangleVec$. With a slight abuse of notation, we denote with $e_{\prec_j}$ the edge in the $j$-th position in $\triangleVec$, and with $\triangleVec(e,\prec)$ the position of edge $e\in E$ in $\triangleVec$. We then define a \emph{perfect} ranking predictor as follows:

\smallskip
(\textsc{Ranking predictor}) Given an integer value $K>0$, a \emph{ranking predictor} $\oracle{\cdot}_K$ is such that $\oracle{e}_K = 1$ iff $\triangleVec(e,\prec) \le K$.
\smallskip

A ranking predictor requires as unique input a parameter $K$, that is, the number of edges to classify as heavy. Clearly, $K$ corresponds to a bound on the maximum number of edges to be retained deterministically by \algName (see \Cref{code:heavy_classification}).\footnote{The set of heavy edges $H$ used by \algName has size trivially bounded by $K$.} Importantly, a ranking predictor does not require the knowledge of a threshold over edge weights $W(e), e\in E$ to classify heavy edges, in contrast with previous literature~\citep{chen2022triangle}. That is our predictor model is required to output the ranking $\triangleVec$ without having explicit access to the weights $W(e),e\in E$, as this would not be practical.

We now obtain a bound on the probability $p$ for which it holds that \algName provides a relative $\varepsilon$-approximation of all the temporal triangle counts $|\triType_i|, i\in[8]$ with controlled error probability and sublinear memory usage. As a simplifying assumption we assume that there exists an arbitrary large constant $C$ for which $|\triType_{i}| = C \cdot  |\triType_{j}|, i,j\in[8] $.\footnote{This assumption is not fundamental and it can be removed in all our results replacing $|\triType_{i}|$ with $\sum_i |\triType_{i}|$.}

\begin{theorem}~\label{theorem:varianceNoNoise}
	Consider an execution of~\algName coupled with a ranking predictor $\oracle{\cdot}_K$ classifying as heavy $K=o(m_\delta)$ edges from $E$ and $\varepsilon>0$ an accuracy parameter. There exist constants $C>1, \gamma\in(0,\tfrac{1}{2})$ such that if $ \ \forall i \in [8]$ it holds $p\ge \tfrac{C\sqrt{ m_\delta}}{\varepsilon|\triType_i|^{1/2-\gamma}}$ then \algName with predictor $\oracle{\cdot}_K$ is a one-pass streaming algorithm with $\Prob[\exists i\in [8] : |\estimate_i -|\triType_i| | \ge \varepsilon |\triType_i|] \le 1/3$, using $\bigO\left( \varepsilon^{-1} m_\delta^{3/2} / |\triType_i|^{1/2-\gamma}\right)$ memory in expectation.
\end{theorem}

The proof is found in~\Cref{appsec:missingProofs}.

Consider the following event $\mathsf{E} = $``the triangle count $|\triType_i|^{1/2-\gamma}$ is sufficiently large, for all eight triangle types''. Then \Cref{theorem:varianceNoNoise} indicates that the sampling probability $p$ can be set as $p \ge \bigO(\varepsilon^{-1} \sqrt{m_\delta}/|\triType_i|^{1/2-\gamma})$, leading to a very small value for $p$, as $\sqrt{m_\delta}/|\triType_i|^{1/2-\gamma}\ll 1$ under condition $\mathsf{E}$. Consequently, the expected memory usage of \algName becomes sublinear in $m_\delta$ under condition $\mathsf{E}$, aligning with established results in concentration theory~\citep{Boucheron2004Concentration}. Clearly, if $\mathsf{E}$ does not hold, then the counts $|\triType_i|$ are small enough so they can be obtained with high accuracy using the set $H$ identified through the predictor $\oracle{\cdot}$.Let now $\rho_K = W(e_{\prec_{K+1}})$ be the (unknown) highest weight of a light edge (according to the predictor) in position $K+1$ in \triangleVec. By our analysis of \Cref{theorem:varianceNoNoise}, we get the following corollary, which captures the reduction in the variance over the estimates $\estimate_i, i\in[8]$ due to the use of a ranking predictor $\oracle{\cdot}_K$.

\begin{corollary}\label{coroll:varianceNoNoise}
Given the setting of \Cref{theorem:varianceNoNoise}, for the estimates $\estimate_i, i\in [8]$ provided in output by \algName  with ranking predictor $\oracle{\cdot}_K$ it holds $ \Var[\estimate_i] \le C p^{-2} \rho_K m_\delta |\triType_i| $. Furthermore, when \algName is executed \emph{without} a predictor $\oracle{\cdot}_K$ it holds that $\Var[\estimate_i] \le C' p^{-2} m_\delta |\triType_i|^2 $ for a sufficiently large constant $C'$.
\end{corollary}
The proof is found in~\Cref{appsec:missingProofs}.

That is, when $\rho_K$ is not too large (i.e., the first $K$ edges in $\triangleVec$ capture most temporal triangle occurrences) then $\Var[\estimate_i] = \bigO(p^{-2} m_\delta |\triType_i|)$ is a factor $\bigO(|\triType_i|)$ sharper than the bound $\Var[\estimate_i] = \bigO(p^{-2} m_\delta |\triType_i|^2)$ that~\algName would achieve \emph{without the ranking predictor}, which is a remarkable variance reduction.

\emph{Noisy predictor.} We now introduce a more practical \emph{noisy} ranking predictor. Again, our definition captures models optimized for ranking metrics, e.g., correlation measures.
Given two parameters $(\alpha, K)$, we denote with $\Pi(\{1,\dots,m\})_{(\alpha, K)}$ the set of permutations of $m$ elements where three blocks of elements are fixed, namely the blocks $[1,\ldots,K-\alpha-1]$,  $[K - \alpha, \ldots, K+\alpha]$ and $[K+\alpha+1, \ldots, \numEdgesTot]$.
That is, elements from one block can only be permuted inside the same block. A \emph{noisy} ranking predictor is defined as follows.

\smallskip
(\textsc{Noisy ranking predictor}) Given a parameter $K>0$ and $0\le\alpha \le \min\{m-K+1, K-1\}$ an \emph{$\alpha$-noisy $K$-ranking predictor} outputs $\triangleVec_\pi=\langle e^\totEdgeW_{\pi_1},\dots, e^\totEdgeW_{\pi_{|E|}}\rangle$, that is the vector $\triangleVec$ permuted according to $\pi \sim \Unif(\Pi(\{1,\dots,\numEdgesTot\})_{(\alpha, K) })$ where $\Unif(\cdot)$ denotes the uniform distribution over the elements of a set.
\smallskip

Therefore an $\alpha$-noisy $K$-ranking predictor is such that it correctly classifies the top-$(K-\alpha-1)$ edges in $\triangleVec$, and the edges with small weight $W(e)$ (i.e., all edges in position $j\in\{K+\alpha+1,\ldots,m\}$ in $\triangleVec$). Instead, the noisy oracle can be arbitrarily wrong in classifying edges in position $K-\alpha,\dots, K+\alpha$ over $\triangleVec$. Hence, $\alpha$ can be viewed as a \emph{noise parameter}, and a larger value for $\alpha$ corresponds to a much less accurate ranking predictor. Our noisy predictor definition closely reflects machine learning models or recommenders with high recall.. In fact, current models achieve high precision over the top-ranked and bottom-ranked elements ($[1,\ldots,K-\alpha-1]$ and $[K+\alpha+1, \ldots, \numEdgesTot]$ blocks in our model, respectively), while they may obtain arbitrary wrong predictions for all other positions ($[K - \alpha, \ldots, K+\alpha]$ block). Note that a $0$-noisy predictor corresponds to the previously defined ranking predictor (without noise).

Let $\boundDiff = \max_{i\in[1,\numEdgesTot-1]} \{\totEdgeWeight{e_{\prec_{i}}} - \totEdgeWeight{e_{\prec_{i+1}}} \}$ be the maximum difference of the weights for two adjacent edges in the vector $\triangleVec$. Finally let $\boundDiff_a = \boundDiff \cdot (a+1), a\ge 0$. We now relate the (unknown) noise parameter $\alpha$ to the guarantees offered by~\algName.

\begin{theorem}\label{theo:noisyGuarantees}
Consider an execution of~\algName coupled with $\oracle{\cdot}_K$, an $\alpha$-noisy $K$-ranking predictor for $\alpha\ge 1$,  $K=o(m_\delta)$ and $\varepsilon>0$. There exist constants $C, \gamma\in (0,1/2)$ such that, if $p\ge \tfrac{C\sqrt{ \boundDiff_\alpha  m_\delta}}{\varepsilon|\triType_i|^{1/2-\gamma}}$, then \algName with predictor $\oracle{\cdot}_K$ is a one-pass streaming algorithm, for which $\Prob[\exists i\in [8] : |\estimate_i -|\triType_i| | \ge \varepsilon |\triType_i|] \le 1/3$, using $\bigO \left(\varepsilon^{-1} m_\delta^{3/2}\sqrt{\boundDiff_\alpha}/|\triType_i|^{1/2-\gamma} \right)$ memory in expectation.
\end{theorem}

The proof is found in~\Cref{appsec:missingProofs}.

Therefore, a noisy predictor increases \algName's variance by a factor $\sqrt{\boundDiff_\alpha}$ (with respect to \Cref{theorem:varianceNoNoise}), where $\boundDiff_\alpha$ reflects the \emph{noise} $\alpha$ over the predictions. In fact, \algName may miss some triangle counts for heavy edges affected by the noisy predictions. Despite this, if the predictor effectively ranks important (heavy) edges (i.e., $\sqrt{\boundDiff_\alpha}$ is not too large), then \algName achieves accurate estimates with reduced variance compared to classical algorithms while maintaining an expected sublinear memory usage.

\subsection{A simple and practical predictor} \label{sec:mdp}

We now introduce a simple and practical ranking predictor, denoted with \emph{temporal min-degree predictor}, to be used within our algorithm~\algName. Our practical predictor is efficiently computable with a single pass on the input stream. 
Formally, consider a node $u \in V$. We define the \emph{temporal degree} of node $u$ within a \emph{time interval} $[t_a, t_b]$ as $\temporalDeg{u,t_a, t_b} = | \{ (x, y, t') \in E : (x = u \text{ or } y = u) \text{ and } t' \in [t_a, t_b] \} |$. That is, the temporal degree is the number of edges incident to $u$ within the given time window. Next, given a temporal edge $e = (u, v, t) \in E$ and a time duration $\delta$ let $\edgeTemporalDeg{e} = \min \{ \temporalDeg{u,t-\delta, t+\delta}, \temporalDeg{v,t-\delta, t+\delta} \}$ be the \emph{temporal min-degree weight}. That is, $\edgeTemporalDeg{e}$ is the minimum between the temporal degrees of the nodes of $e$. Intuitively, the temporal min-degree weight captures the temporal activity of individual nodes over the graph, which is crucial for our goal of designing a highly accurate and practical predictor for~\algName. Let $\triangleVec^{\texttt{m-d}}$ be the edges $e\in E$ ordered by non-increasing values of their weight $\edgeTemporalDeg{e}$, ties broken arbitrarily. Then, for any edge $e \in E$, the \emph{temporal min-degree ranking predictor} classifies $\oracle{e}_K = 1$ if $e$ is within the first $K$ edges of $ \triangleVec^{\texttt{m-d}}$, and $\oracle{e}_K = 0$ otherwise. Therefore, the temporal min-degree predictor uses the temporal min-degree weights $\edgeTemporalDeg{e}$ as a \emph{proxy} for the unknown values $W(e)$ of a perfect predictor. Clearly, the predictions provided by the temporal min-degree predictor may not be accurate when the rankings of $\triangleVec^{\texttt{m-d}}$ and $\triangleVec$ do not align, see \Cref{theo:noisyGuarantees}. Note that the temporal min-degree predictor can be computed extremely efficiently, with a single pass over the stream, and avoiding exact temporal triangle counting. Finally, note that our temporal min-degree predictor leverages both \emph{structural} and \emph{temporal} properties in the data,
making it significantly different from predictors for static triangle counting~\citep{chen2022triangle,boldrin2024fast}, that do not consider time. In \Cref{sec:ablation}, we provide an empirical comparison for~\algName coupled with our temporal min-degree predictor and two predictors based on state-of-the-art algorithms for \emph{static} triangle counting, showing the superior performance of our temporal min-degree predictor.

\section{Experimental evaluation}\label{sec:experiments}

We present our extensive experimental evaluation, focusing on the following questions: 

\textbf{Q1}. How does \algName compare to \sota approaches in terms of accuracy of its estimates and computational resources (time and memory)? \label{exp:Q1} 

\textbf{Q2}. What is the impact of the predictor $\oracle{\cdot}$ on the estimates of \algName? \label{exp:Q2} 

\textbf{Q3}. How does \algName perform in an \emph{online setting}, where the predictor $\oracle{\cdot}$ is learned on historical data and then used on previously unseen data? \label{exp:Q3} 

\textbf{Q4.} Can we improve our min-degree predictor that is designed specifically for \Cref{prob:ourProb} by using  \emph{static} information? \label{exp:Q4} 

\para{Datasets.} We considered four massive publicly available temporal graphs (\cref{tab:datasets}), on which previous studies~\citep{pan2024accurate,gao2022scalable,Sarpe2024TMDS,pashanasangi2021faster} have shown that solving~\cref{prob:ourProb} is extremely challenging. 
Briefly, \emph{Stackoverflow} (SO) \citep{paranjape2017motifs} collects timestamps interactions between users of the corresponding website; \emph{Bitcoin} (BI) \citep{liu2019sampling,Kondor-2014-bitcoin} represents Bitcoin transactions between users; \emph{Reddit} (RE) \citep{liu2019sampling,hessel2016science} is a temporal graph of Reddit comments interactions; \emph{EquinixChicago} (EC), with more than 3 billion edges, is a semi-synthetic graph that we built from the bipartite graph in~\citep{liu2019sampling,sarpe2021presto}. See \cref{appendix::equinix_and_preproc} for more details.

\para{Implementation and setup.} We implemented \algName in C++17, the code was compiled under gcc 9.4.0 with optimization flags enabled. We performed all the experiments on a system running Ubuntu 20.04, with a 2.20 GHz Intel Xeon CPU, and a limited maximum RAM memory of 200GB. The code is available for review and will be released upon acceptance.\footnote{\url{https://github.com/VandinLab/STEP}}

    \begin{table}[t]
    	\centering
    	%\captionsetup{width=\textwidth}
    	\caption{Datasets. We report: $n = |V|$ the number of nodes; $m=|E| $ the number of temporal edges;  the $precision$ of the timestamps; and the total $timespan$.}
    	\label{tab:datasets}
	%\resizebox{\columnwidth}{!}{
    		\begin{tabular}{lcccr}
    			\toprule
    			\bf{Dataset} & \bf{$n$} & \bf{$m$} & \bf{$precision$} & \bf{$timespan$} \\
    			\midrule
    			Stackoverflow (SO) &  $2.6\,M$  & $63.5\,M$ & sec & $7.60$ years\\
    			Bitcoin (BI) & $48.1\,M$  & $113.1\,M$ & sec & $7.08$ years\\
    			Reddit (RE) & $8.4\,M$ &$636.3\,M$ & sec & $10.06$ years\\
    			EquinixChicago (EC) & $11.1\,M$ & $3.3\,B$ & ${\mu}$-sec &  $62.00$ mins\\
    			\bottomrule
    		\end{tabular}
		%}
		%\vspace{-0.4cm}
    \end{table}

\para{Baseline methods.} We compared \algName with the following \sota algorithms: \Degen \citep{pashanasangi2021faster}, an \emph{exact} algorithm for computing the counts of all temporal triangles; \FastTri~\citep{gao2022scalable}, an \emph{exact} algorithm specifically tailored to temporal triangles; \Motto~\citep{li2024motto}, a recent \sota \emph{exact} algorithm for counting 3-nodes 3-edges motifs, including triangles; and \EWS~\citep{wang2022efficient}: the \sota \emph{approximate} method for solving \Cref{prob:ourProb}. We set the parameters of \EWS as $\EWSp=0.01$ and $\EWSq=0.1$, for all datasets, as suggested by the authors~\citep{wang2022efficient} (see \Cref{appendix::parameters} for further details). We also compared with the sampling approach adopted by \algName but without leveraging a predictor, which we denote with \NaiveS. We note that the exact approaches \Degen, \FastTri, and \Motto cannot process the input data in a streaming fashion.

\para{Metrics and parameters.}
\emph{Accuracy.} We consider the Mean Absolute Error (MAE) over ten runs and its standard deviation as metrics for the accuracy of approximate algorithms. Where the MAE corresponds to the estimation error $|\estimate_i - |\triType_i|| / |\triType_i|$ averaged over ten independent runs for each $T_i, i\in[8]$. All the exact counts $|\triType_i|, i \in [8]$ are obtained using a na\"ive enumeration algorithm.

\emph{Memory and runtime.} We measured the peak RAM memory (in GB) required by each algorithm over a representative run. The runtime we report is an average over ten runs, unless otherwise stated. Since \EWS counts triangles independently, its runtime is the aggregate time to estimate all temporal triangle counts, averaged across ten runs.

\emph{Parameters.} We select a small, a medium, and a large value for the parameter $\delta$ according to the precision of each dataset. That is, we set on SO, BI and RE $\delta \in \{3\,600, 86\,400, 259\,200\}$, while for EC we set $\delta \in \{1 \times 10^5, 2 \times 10^5, 3 \times 10^5\}$. We set the parameter $K$ to $\tfrac{m}{100}$. The sampling probability \NaiveSp of the \NaiveS algorithm is set to obtain, in \emph{expectation}, the same number of edges retained by \algName (i.e., $|S_L| + |H|$). All parameters used in our experiments are in \Cref{appendix::parameters}.

\para{Ranking predictors.} We considered two ranking predictors for \algName: a \emph{perfect (impractical) predictor} and a practical \emph{temporal min-degree} predictor: 
\emph{1)} The \emph{perfect predictor} exactly classifies the $K$ edges with the highest weights $\totEdgeW(e), e \in E$ as defined in \Cref{subsec:oraclesAndVariance}. 
\emph{2)} The \emph{temporal min-degree} predictor described in \Cref{sec:mdp}. 
We denote the resulting methods with \algNameWPerfect and \algNameWTemporal respectively. Note that the perfect predictor cannot be leveraged in practical applications, but it allows to evaluate the performance of \algName when predictions are \emph{perfect}, providing a lower bound on the error of the estimates of~\algName. The temporal min-degree predictor is instead simple, general, and domain-agnostic. Given that it can be computed from simple structural properties in the data, the resulting method, \algNameWTemporal, can be easily used in \emph{practice}.

\begin{figure}
	\centering
	\includegraphics[width=0.8\textwidth]{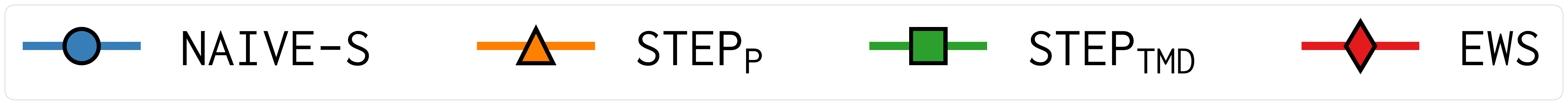}
	\includegraphics[width=0.9\textwidth]{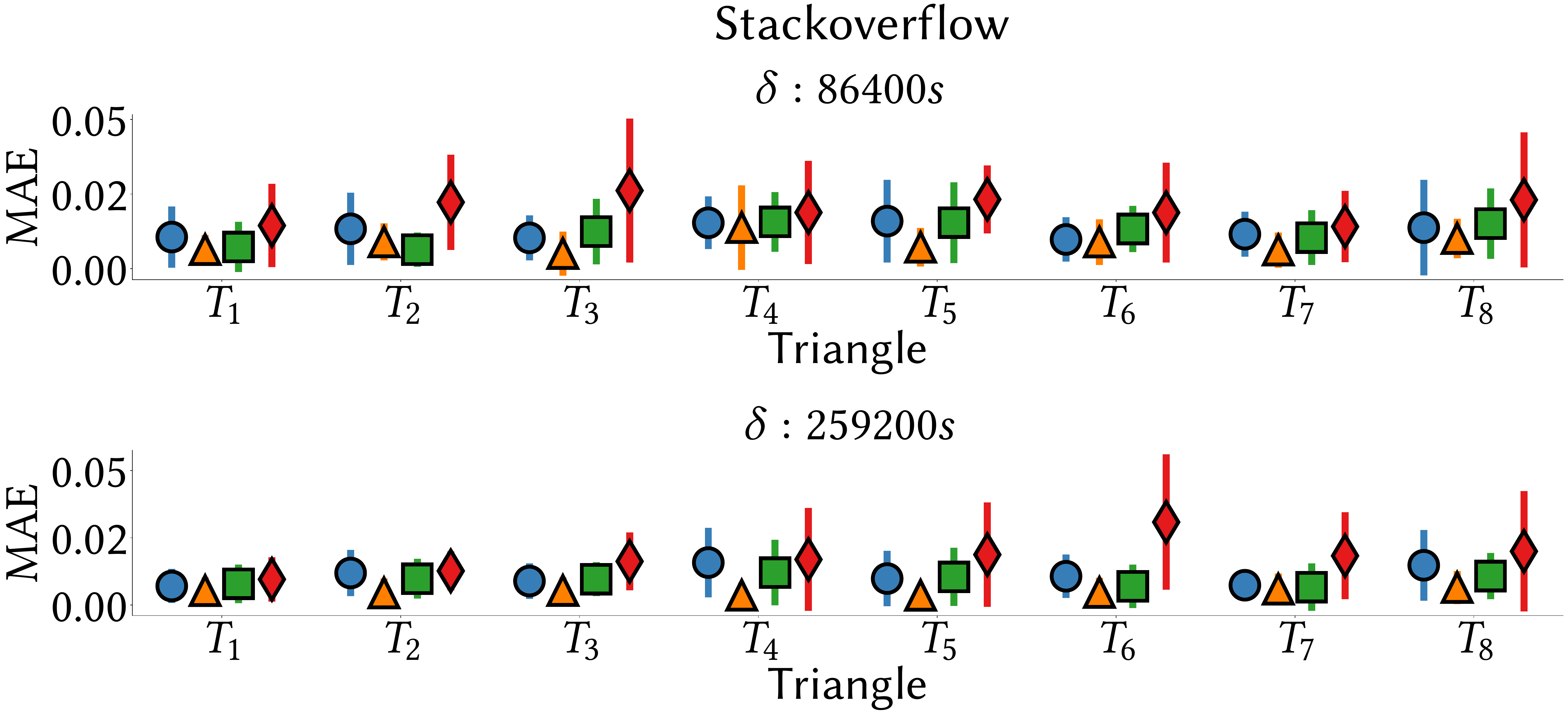}
	\includegraphics[width=0.9\textwidth]{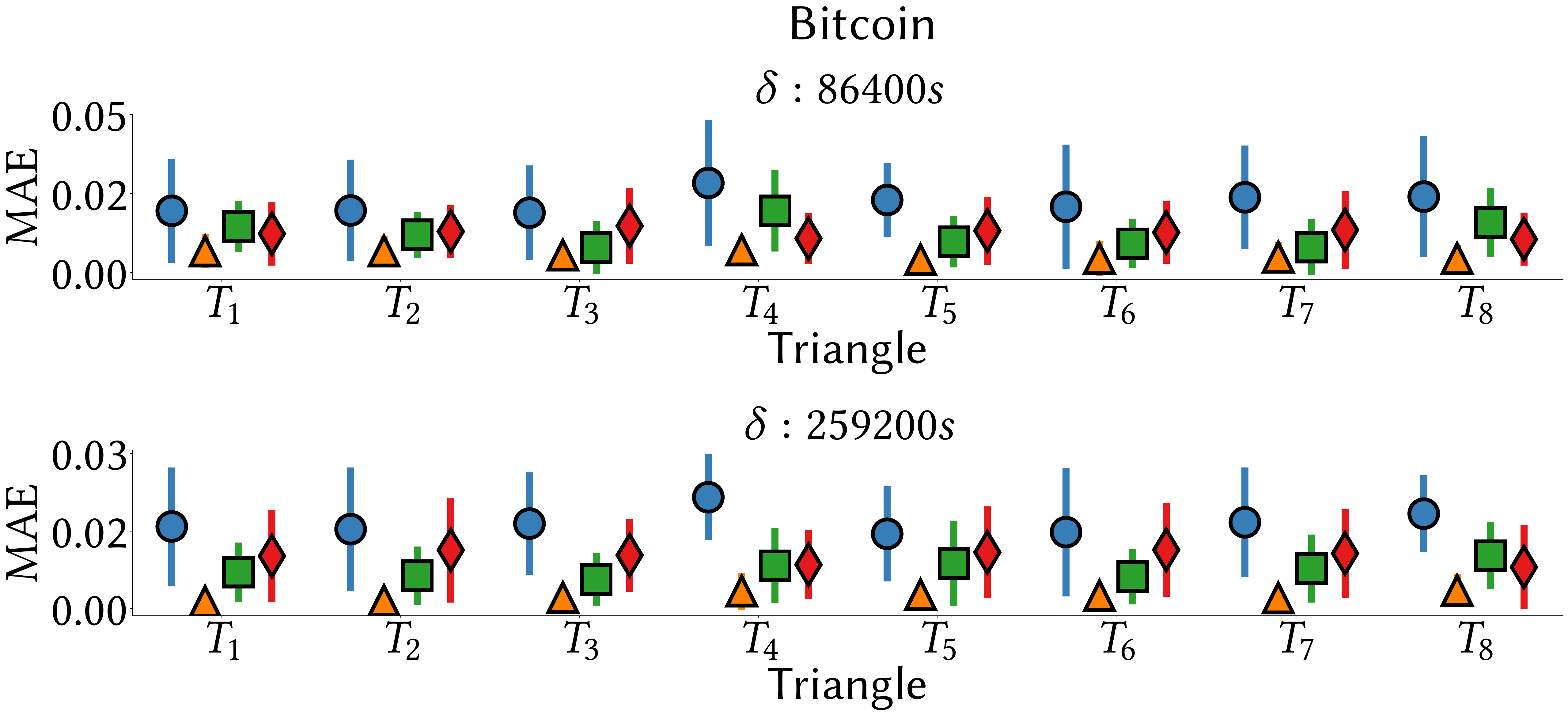} 
	\includegraphics[width=0.9\textwidth]{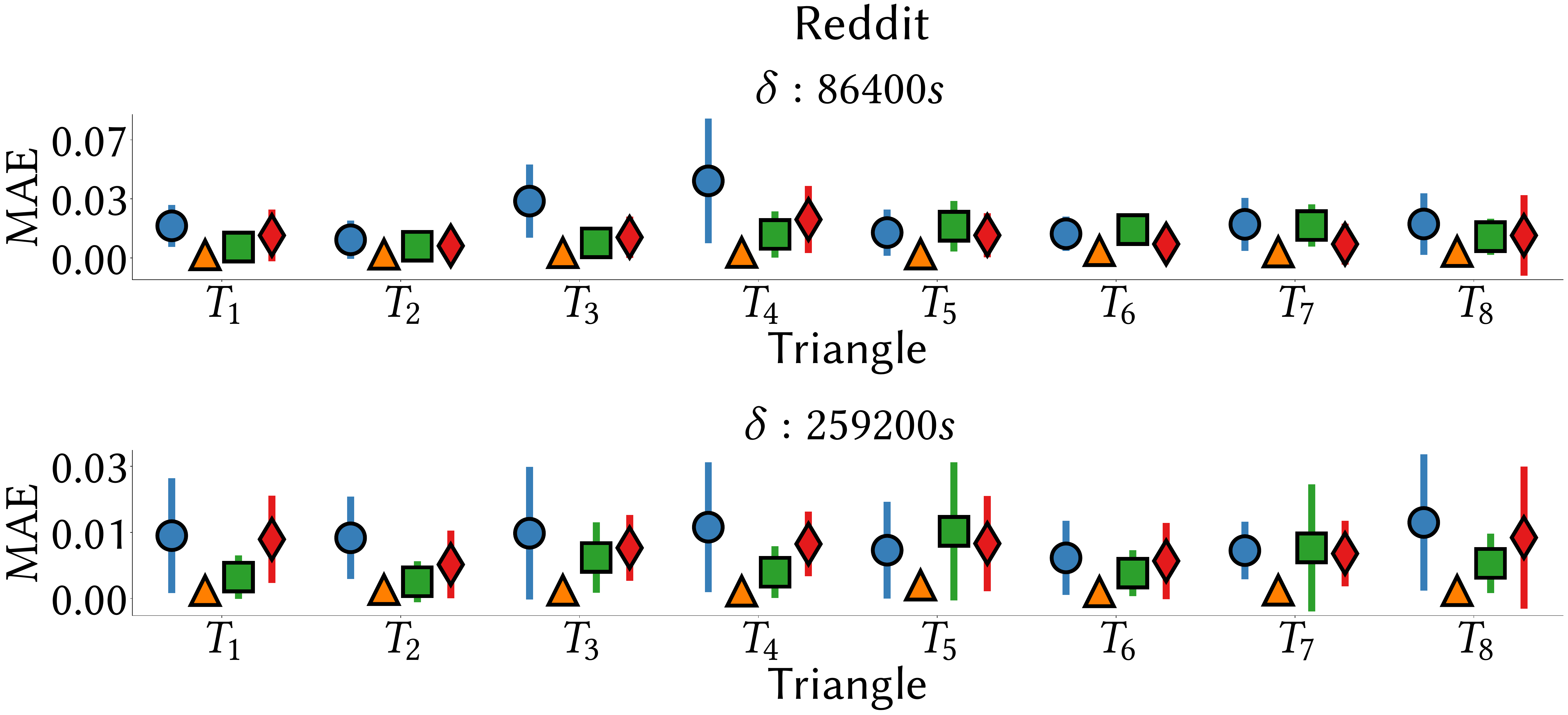}
	\caption{MAE and standard deviation for \algName, \NaiveS and \EWS on SO, BI and RE datasets from \Cref{tab:datasets}, for the largest values of $\delta$ and for each temporal triangle (see \Cref{fig:triangles}(d)).}
        \label{fig:error_results}
\end{figure}

\begin{figure}
	\centering
	\includegraphics[width=0.6\textwidth]{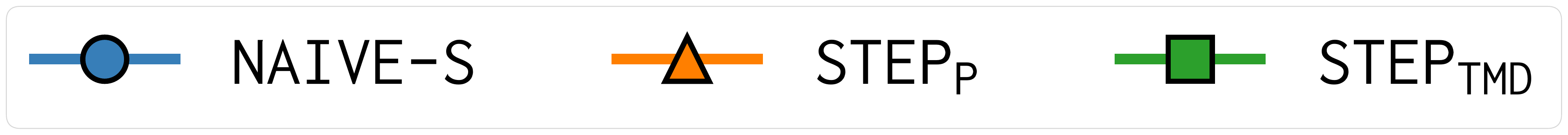}
	\includegraphics[width=0.9\textwidth]{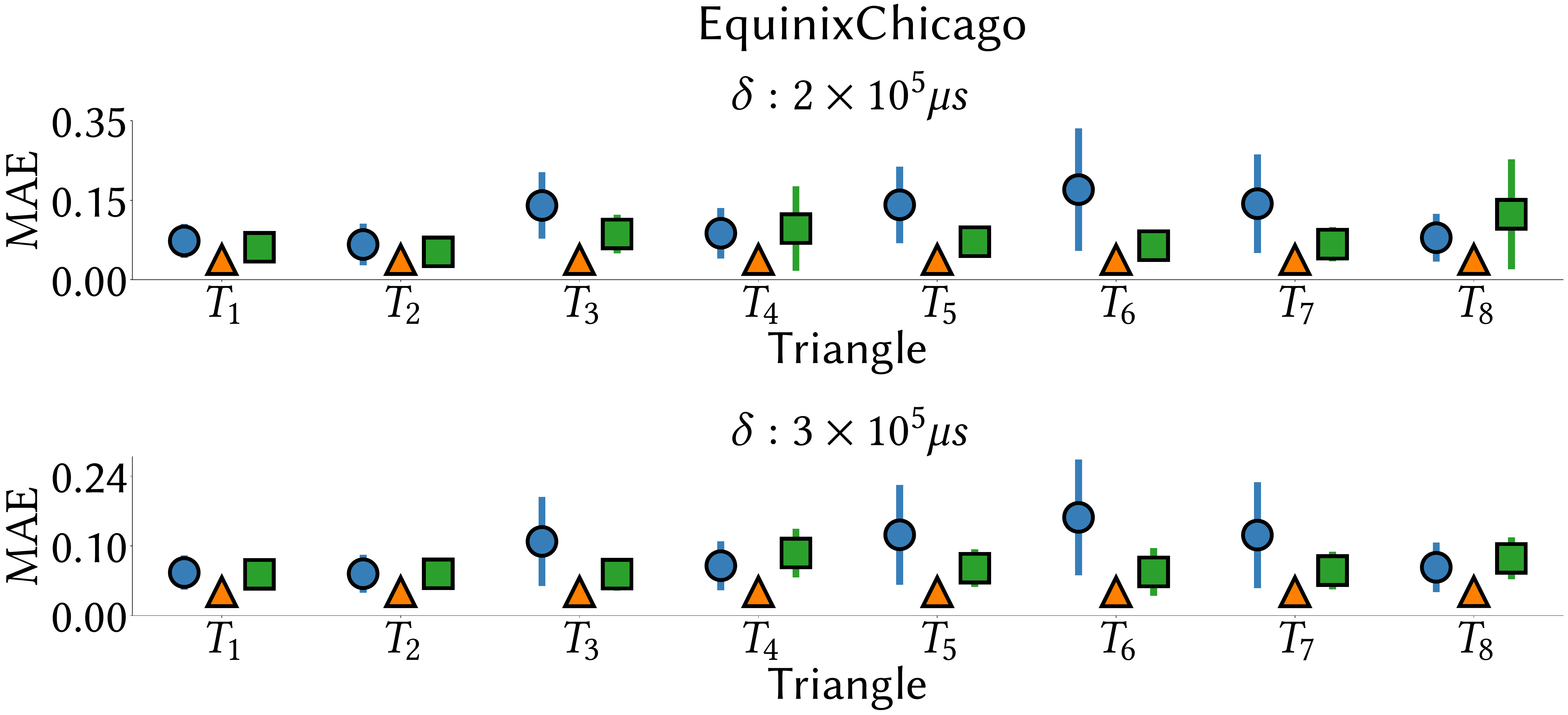}
	\caption{MAE and standard deviation for \algName, \NaiveS and \EWS on EC dataset from \Cref{tab:datasets}, for the largest values of $\delta$ and for each temporal triangle (see \Cref{fig:triangles}(d)).}
	\label{fig:error_results_2}
\end{figure}

\begin{table}[t]
	%\centering
	%\captionsetup{width=\textwidth}
	\caption{Peak RAM memory, in GB, of a representative run for the largest $\delta$. $OOM$ denotes out of memory.}
	\label{tab:mem_results}
    \resizebox{\columnwidth}{!}{
	\begin{tabular}{lccccccc}
	\toprule
	\bf{Dataset} &
	\NaiveS&
	$\algNameWPerfect$ &
	$\algNameWTemporal$ &
	\EWS &
	\Degen &
	\FastTri &
	\Motto \\
	\midrule
	SO & $0.78$ & $0.74$ & $0.74$ & $8.04$ & $5.10$ & $5.81$ & $12.07$\\
	BI & $1.70$ & $1.81$ & $1.74$ & $27.05$ & $15.40$ & $17.43$ & $34.08$\\
	RE & $14.68$ & $14.74$ & $15.53$ & $103.75$ & $71.30$ & $79.59$ & $159.74$ \\
	EC & $63.46$ & $62.70$ & $63.47$ & $OOM$ & $OOM$ & $OOM$ & $OOM$ \\
	\bottomrule
	\end{tabular}
    }
\end{table}

\subsection{Comparison with state-of-the-art methods} \label{sec:soa_comp}

We first compared \algName with \sota baselines to answer question \textbf{Q1}. For such comparison, we fixed the parameters of all the approximation algorithms (\algName, \NaiveS, and \EWS) to ensure comparable runtime and MAE. Specifically, we ran \algName with the same sampling probability as \EWS ($p = 0.01$) on all datasets except for SO, where $p=0.1$ was used since \algName is much faster than \EWS. When discussing results for \algName, we focus on \algNameWTemporal. \Cref{tab:mem_results} shows the peak memory usage for each algorithm, and \Cref{tab:time_results_small} presents the runtime of all algorithms except \algNameWPerfect and \NaiveS (see \Cref{appendix:additional_results} for additional results). \Cref{fig:error_results} depicts the accuracy of each approximate method for the largest values of $\delta$ (exact methods always report 0 MAE). Results for other values of $\delta$ are available in \Cref{appendix:additional_results}. On the SO dataset, \algNameWTemporal requires substantially less memory than \EWS, \Degen, \FastTri and \Motto while achieving more accurate estimates than \EWS for all values of $\delta$ and for most temporal triangle counts. On the RE dataset, \algNameWTemporal similarly demonstrates a significant reduction in memory usage compared to \EWS, \Degen, \FastTri and \Motto, and often provides higher-quality estimates than \EWS. On the BI dataset, \algNameWTemporal is much more memory efficient compared to \EWS, \Degen, \FastTri and \Motto. For larger values of $\delta$, when the estimation problem becomes more challenging, \algNameWTemporal obtains more accurate estimates than \EWS. For smaller values of $\delta$, the estimates provided by \algNameWTemporal are comparable but slightly less accurate than those of \EWS. In terms of runtime, \algNameWTemporal consistently outperforms \Degen, \FastTri and \Motto, and, in most cases, also \EWS, with the exception of $\delta = 259\,200$ on the BI dataset. 
This exception is given by the fact that the BI dataset contains almost 40 billion $\delta$-instances---most of which are deterministically counted by \algNameWTemporal yielding very accurate estimates at the expense of high execution times. See \Cref{appendix::parameters} for further analyses on such important time-accuracy trade-off. 
Finally, on the EC dataset, that has more than 3 billion temporal edges, all \sota baselines cannot terminate their execution within the maximum memory allowance (200GB), whereas \algNameWTemporal requires less than 65GB of memory. In contrast, \algNameWTemporal achieves an average MAE below 0.1 and runs in less than ten minutes, even for the largest $\delta$. This highlights that \algNameWTemporal is a highly efficient algorithm, particularly on massive temporal graphs where \sota approaches cannot scale their computation. 

In summary, \algNameWTemporal enables the \emph{efficient} and \emph{accurate} estimation of all temporal triangle counts with remarkably \emph{small memory usage}, especially on massive datasets where existing methods cannot scale their computation.

\begin{table}[t]
	\centering
	\caption{Average runtime (in seconds). ``\xmark'' denotes out of RAM memory (200 GB). We executed exact algorithm once due to their high runtime, hence we do not display their variance. The best runtime is in bold. \texttt{SU} denotes the speed-up of \algNameWTemporal compared to each baseline (\texttt{N/A} is used when the speed-up cannot be computed).}
	\label{tab:time_results_small}
	\resizebox{\textwidth}{!}{
	\begin{tabular}{llccccccccc}
		\toprule
		Dataset &
		$\delta$ &
		$\algNameWTemporal$ &
		\multicolumn{2}{c}{\EWS} &
		\multicolumn{2}{c}{\Degen} &
		\multicolumn{2}{c}{\FastTri} &
		\multicolumn{2}{c}{\Motto} \\
		\hline
		& & \texttt{Time} & \texttt{Time} & \texttt{SU} & \texttt{Time} & \texttt{SU} & \texttt{Time} & \texttt{SU} & \texttt{Time} & \texttt{SU} \\
		\midrule
		\multirow{3}{*}{SO} & $3600$ & \textbf{4.9 $\pm$ 0.0} & $30.4 \pm 0.8$ & $\mathtt{6.2\times}$ & $348.4$ & $\mathtt{71.1\times}$ & $15.1$ & $\mathtt{3.1\times}$ & $174.5$ & $\mathtt{35.6\times}$ \\
		& $86400$ & \textbf{6.3 $\pm$ 0.1} & $32.0 \pm 0.5$ & $\mathtt{5.1\times}$ & $355.2$ & $\mathtt{56.4\times}$ & $41.8$ & $\mathtt{6.6\times}$ & $254.5$ & $\mathtt{40.4\times}$ \\
		& $259200$ & \textbf{7.5 $\pm$ 0.1} & $35.6 \pm 0.9$ & $\mathtt{4.7\times}$ & $356.3$ & $\mathtt{47.5\times}$ & $76.3$ & $\mathtt{10.2\times}$ & $378.9$ & $\mathtt{50.5\times}$ \\
		\midrule
		\multirow{3}{*}{BI} & $3600$ & \textbf{6.1 $\pm$ 0.1} & $67.7 \pm 5.1$ & $\mathtt{11.1\times}$ & $422.1$ & $\mathtt{69.2\times}$ & $189.0$ & $\mathtt{31.0\times}$ & $1113.7$ & $\mathtt{182.6\times}$ \\
		& $86400$ & \textbf{43.8 $\pm$ 0.4} & $115.9 \pm 1.7$ & $\mathtt{2.6\times}$ & $421.3$ & $\mathtt{9.6\times}$ & $4287.7$ & $\mathtt{97.9\times}$ & $18045.1$ & $\mathtt{412.0\times}$ \\ 
		& $259200$ & $278.2 \pm 5.5$ & \textbf{198.5 $\pm$ 5.9} & $\mathtt{0.7\times}$ & $424.8$ & $\mathtt{1.5\times}$ & $13804.3$ & $\mathtt{49.6\times}$ & $55494.2$ & $\mathtt{199.5\times}$ \\
		\midrule
		\multirow{3}{*}{RE} & $3600$ &  \textbf{69.7 $\pm$ 2.0} & $570.7 \pm 50.3$ & $\mathtt{8.2\times}$ & $18656.8$ & $\mathtt{267.7\times}$ & $1708.7$ & $\mathtt{24.5\times}$ & $6406.0$ & $\mathtt{92.0\times}$ \\
		& $86400$ & \textbf{103.5 $\pm$ 4.4} & $943.1 \pm 67.6$ & $\mathtt{9.1\times}$ & $18528.1$ & $\mathtt{179.0\times}$ & $7479.5$ & $\mathtt{72.3\times}$ & $23085.0$ & $\mathtt{223.0\times}$ \\ 
		& $259200$ & \textbf{164.9 $\pm$ 2.1} & $1121.8 \pm 79.1$ & $\mathtt{6.8\times}$ & $18950.0$ & $\mathtt{115.0\times}$ & $11165.8$ & $\mathtt{67.7\times}$ & $29680.1$ & $\mathtt{180.0\times}$ \\
		\midrule
		\multirow{3}{*}{EC} & $1 \times 10^5$ & \textbf{425.6 $\pm$ 16.8} & \xmark & \texttt{N/A} & \xmark & \texttt{N/A} & \xmark & \texttt{N/A} & \xmark & \texttt{N/A} \\
		& $2 \times 10^5$ & \textbf{497.4 $\pm$ 8.6} & \xmark & \texttt{N/A} & \xmark & \texttt{N/A} & \xmark & \texttt{N/A} & \xmark & \texttt{N/A} \\
		& $3 \times 10^5$ & \textbf{580.3 $\pm$ 0.9} & \xmark & \texttt{N/A} & \xmark & \texttt{N/A} & \xmark & \texttt{N/A} & \xmark & \texttt{N/A} \\
		\bottomrule
	\end{tabular}
}
\end{table}

\subsection{Impact of the predictor} \label{sec:pred_impact}

To answer \textbf{Q2}, we consider \algNameWPerfect, \algNameWTemporal, and \NaiveS, and set their parameters so that they sample the same number of edges in expectation (see \Cref{appendix::parameters}). \Cref{fig:error_results} shows that both \algNameWPerfect and \algNameWTemporal provide much more accurate estimates than \NaiveS on all the datasets, with the exception of  $\delta=3\,600$ for the BI dataset where \algNameWTemporal and \NaiveS are comparable (see \Cref{appendix:additional_results}). Moreover, the variance of the estimates by both \algNameWPerfect and \algNameWTemporal is always smaller compared to \NaiveS, especially for the larger datasets RE and EC. In terms of runtime, \NaiveS is the fastest one (see \Cref{tab:time_results} in \Cref{appendix:additional_results}): this is because the \NaiveS \emph{counts} fewer triangles than \algName, yielding estimates with higher variance. This is an unavoidable trade-off, that is, more accurate estimates require larger execution times for \algName. It is worth noting that on the EC dataset, \algNameWTemporal takes more time to execute than \algNameWPerfect, in contrast to all other datasets. This is due to the structure of the EC dataset (i.e., starting from a bipartite graph, see \Cref{appendix::equinix_and_preproc}), on which the temporal min-degree does not provide a good proxy for the weights $W(e)$ of temporal edges (see \Cref{appendix::predictors_correlation} for more details). Nevertheless, \algName still computes better estimates than \NaiveS while being highly memory efficient. 

To summarize, the experiments show that, by employing a predictor, \algNameWPerfect and \algNameWTemporal significantly improve the accuracy and reduce the variance of the estimates of \NaiveS, while having slightly higher execution times. In fact, \NaiveS identifies fewer triangles than \algNameWPerfect and \algNameWTemporal, being faster but significantly less accurate.

\begin{figure}[H]
	\centering
	\includegraphics[width=0.4\textwidth]{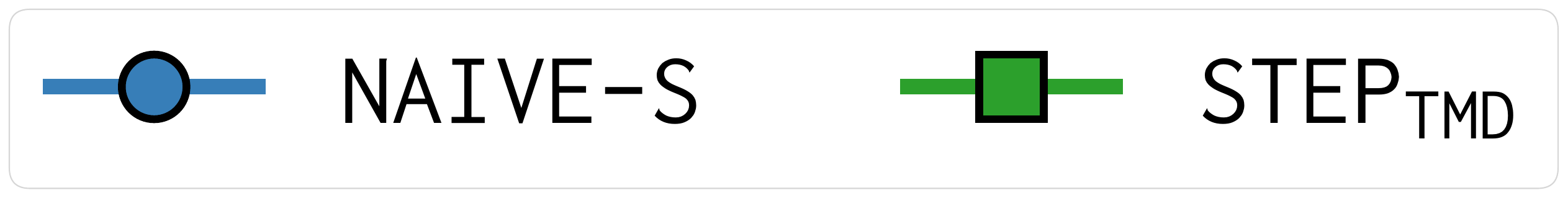}
	\includegraphics[width=0.8\textwidth]{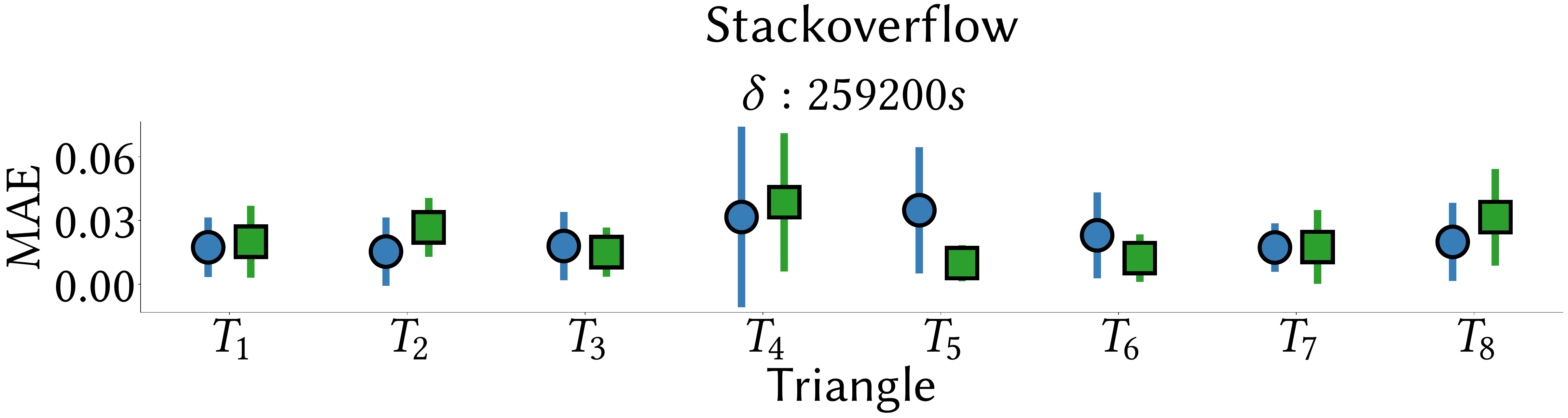} 
	\includegraphics[width=0.8\textwidth]{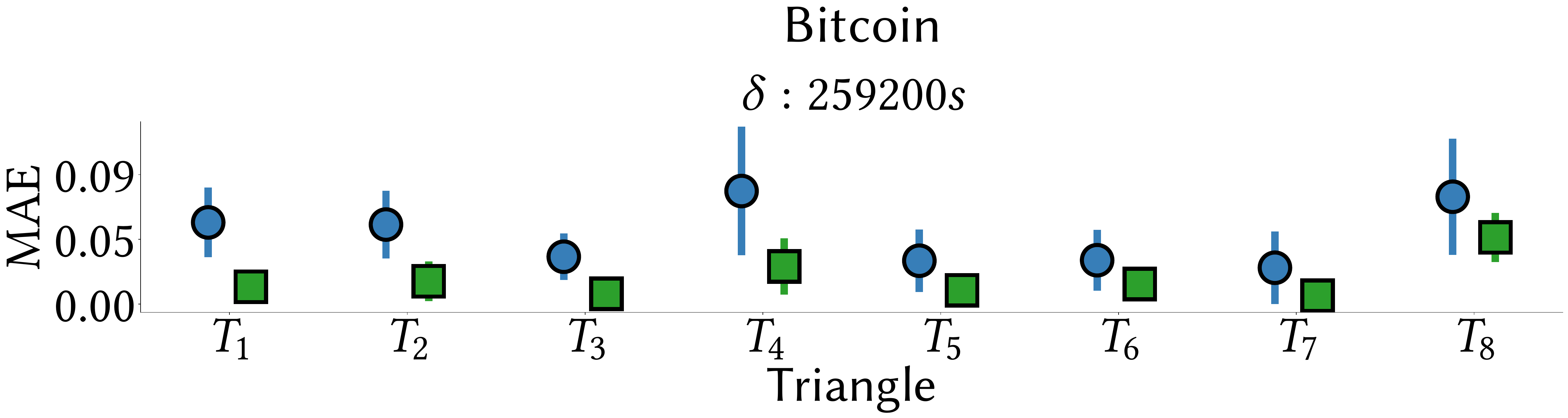} 
	\includegraphics[width=0.8\textwidth]{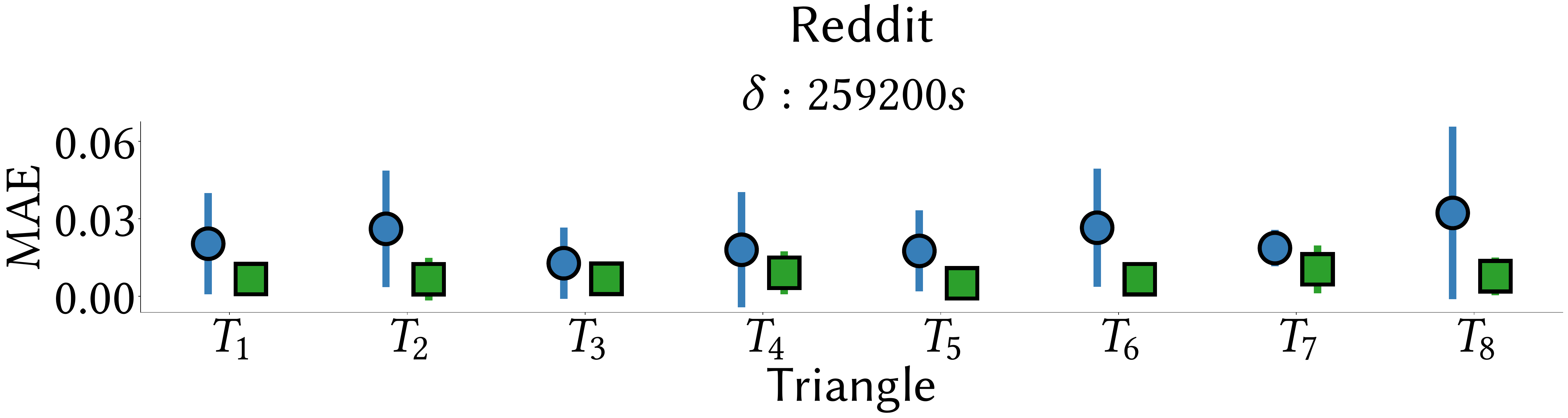}
	\includegraphics[width=0.8\textwidth]{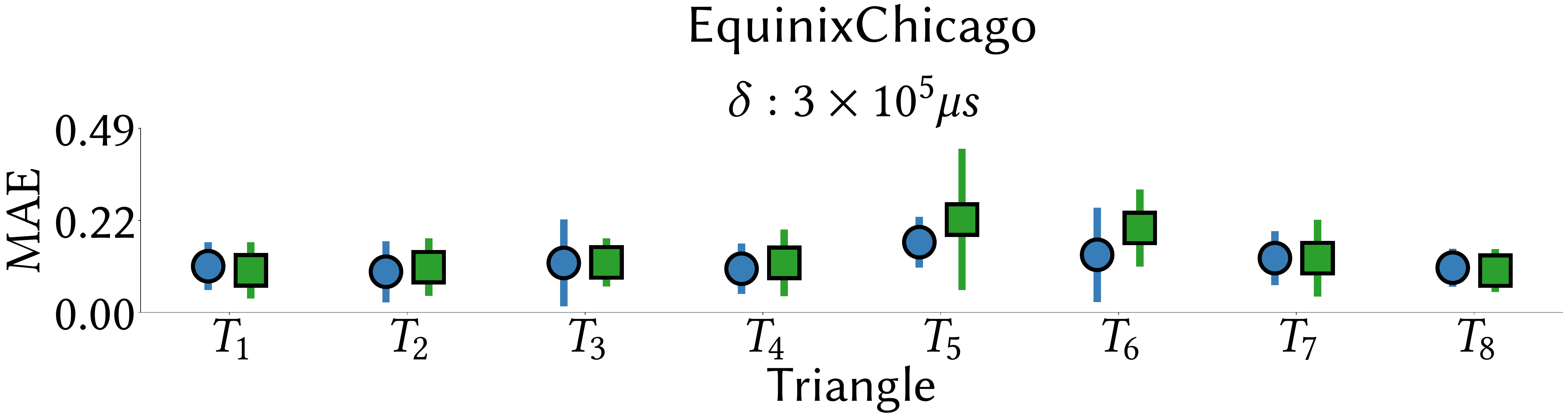}
	\caption{MAE and standard deviation for online estimation on the stream $\tau^{ts}$ by the \NaiveS algorithm and \algNameWTemporal (trained on the historical data from $\tau^{tr}$) on each dataset, for the largest value of $\delta$ and for each triangle type (see \Cref{fig:triangles}(d)).}
	\label{fig:error_results_split_stream}
\end{figure}

\subsection{Online estimation} \label{sec:online_est}

To address \textbf{Q3}, we developed a practical approach for learning a predictor from a \emph{training stream} of temporal edges ($\tau^{tr}$) and use it to estimate temporal triangle counts on a \emph{test stream} ($\tau^{ts}$). The training stream $\tau^{tr}$ consists of the first 75\% of edges appearing in the stream of a graph. The predictor is based on a threshold value $\predThreshold$, derived from the temporal min-degree weight (see \Cref{sec:mdp}) of the $K$-th edge in the non-decreasing ordering induced by $\edgeTemporalDeg{e}, e\in \tau^{tr}$. When processing the test stream, edges in $\tau^{ts}$ with temporal degree $\edgeTemporalDeg{e} \ge \predThreshold$ are classified as heavy and retained for further computation.\footnote{Note that, the predictor evaluates if $e = (u,v,t)$ should be retained or not at time $t + \delta$.} The underlying idea is that the threshold $\predThreshold$ can be used to detect important edges to retain over $\tau^{ts}$ whenever the training stream $\tau^{tr}$ is sufficiently representative. The results for the largest value of $\delta$ are shown in \Cref{fig:error_results_split_stream} (complete results are in \Cref{appendix:additional_results}). We observe that \algNameWTemporal provides more accurate estimates than \NaiveS on the RE and BI datasets, and comparable or slightly better performance on SO. On the EC dataset, \algNameWTemporal and \NaiveS achieve similar accuracy, as the learned predictor does not align well with a \emph{perfect} classification over $\tau^{ts}$ (similarly to the results in \Cref{sec:pred_impact}). Further results are in \Cref{appendix::predictors_correlation}. 

Overall, our findings highlight that \algName effectively leverages our simple predictors learned from historical data, often outperforming \NaiveS in most configurations. 
Therefore, even under very noisy predictions, \algName achieves accurate or satisfactory quality for its estimates. This, supports our design of~\algName and motivates the use of~\algName for practical and challenging applications such as the online processing of temporal graphs.

\begin{figure}
	\centering
	\includegraphics[width=\textwidth]{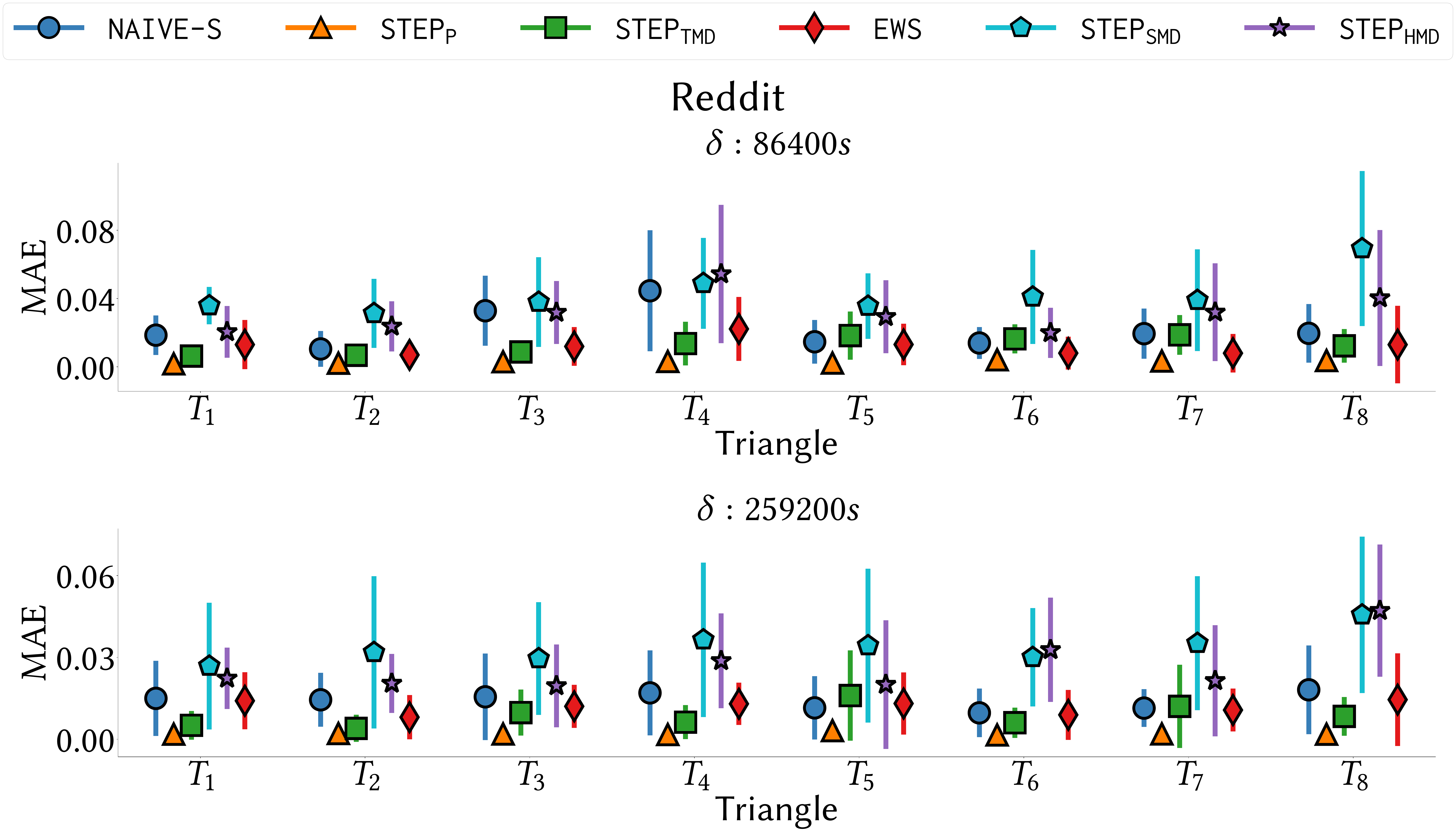}\\
	\caption{MAE and standard deviation for \algName (with different predictors) and \NaiveS on the RE dataset, for the largest values of $\delta$ considered and for each temporal triangle (see \Cref{fig:triangles}(d)).}
	\label{fig:error_ablation}
\end{figure}

\subsection{Ablation study} \label{sec:ablation}

To answer question \textbf{Q4}, we conducted an ablation study on our practical \emph{temporal min-degree} predictor by comparing it with naive predictors based on static node degrees (e.g., the one from~\citep{boldrin2024fast}). 
We considered: $i$) a \emph{static} predictor (denoted with \algNameWStatic when coupled with \algName) which computes the weight of a temporal edge $(u,v,t)$ as $w_{\mathtt{st}}(e) = \min \{ \temporalDeg{u}, \temporalDeg{v} \}$ where $\temporalDeg{\cdot}$ denotes the \emph{static} degree of a node, i.e., $d(u) = |\{ v \in V: v \text{ is connected to } u \}|$; 
$ii$) an \emph{hybrid} predictor (denoted with \algNameWHybrid when coupled with \algName) that ignores the multiplicity of temporal edges between a given pair of nodes, resulting in an edge weight given by $w_{\mathtt{hmd}}(e)= \min \{ \tilde{d}(u, t - \delta, t + \delta), \tilde{d}(v, t - \delta, t + \delta) \}$ where $\tilde{d}(u, t_a, t_b) = | \{  v \in V : v \text{ is connected to } u \text{ within the time window } [t_a, t_b] \} |$. \Cref{fig:error_ablation} shows that \algNameWStatic and \algNameWHybrid, achieve estimates with high MAE and variance, much worse than \algNameWTemporal. 
\Cref{fig:error_ablation} confirms that our temporal min-degree predictor effectively captures important temporal edges to be retained by~\algName. 

Summarizing, embedding static information as done in existing \sota static predictors fails when applied to our temporal scenario~\citep{boldrin2024fast}, motivating our design of a novel predictor. Additional results on the effectiveness of our temporal min-degree predictor are in \Cref{appendix::predictors_correlation}.

\section{Related work}\label{sec:related_works}

To the best of our knowledge, our work is the first to solve the problem of counting \emph{temporal} triangles \emph{with predictions}, that is, using the \awp framework~\citep{mitzenmacher2022algorithms}.
We now discuss relevant works, we defer the reader interested to temporal motifs to~\citep{Liu2021MotifsDefs,Gionis2024Tutorial} and algorithms with predictions to~\citep{mitzenmacher2022algorithms}.

\para{Temporal motif and triangle counting.} 
There exist several definitions of temporal motifs, including temporal triangles~\citep{Liu2021MotifsDefs,kumar20182scent,mang2024efficient}. We consider the widely adopted definition introduced by \citet{paranjape2017motifs}, used in a variety of practical applications. For such definition, various exact and approximate counting methods exist.

\emph{Exact methods.} 
\cite{paranjape2017motifs} introduced dynamic programming algorithms with a complexity of $\bigO(|E'|^{3/2} + m |E'|^{3/4})$, where $|E'|$ is the number of edges in the static graph obtained from a temporal graph, which is impractical for large temporal graphs. \cite{gao2022scalable} and \cite{li2024motto} further improved the work in~\citep{paranjape2017motifs} designing algorithms with various pruning techniques yielding a time complexity of $\bigO(m m_\delta^2)$, matching the complexity of the algorithm in~\citep{Mackey2018EdgeDriven}. However, both most recent works \citep{li2024motto} and \citep{gao2022scalable} do not scale to large temporal graphs, as shown by our experimental evaluation (\Cref{sec:experiments}). \cite{pashanasangi2021faster} developed an exact method with complexity $\bigO(m\kappa \log m)$, where $\kappa$ is the degeneracy of the underlying static graph~\citep{matula1983smallest}. Such an approach can be impractical on large real-world graphs where $\kappa$ is in the order of hundreds or thousands~\citep{pashanasangi2021faster}. Moreover, the approaches mentioned above~\citep{gao2022scalable, pashanasangi2021faster,li2024motto} remain computationally demanding and memory-intensive, as they need full access to the temporal graph.

\emph{Approximate methods.} Approximate approaches are mostly based on randomized sampling. Various such methods exist to approximate a \emph{single} temporal motif count, either by collecting subgraphs within specific time windows~\citep{sarpe2021presto,liu2019sampling} or by sampling temporal edges~\citep{wang2022efficient} or paths~\citep{pan2024accurate}. Some techniques can also estimate multiple temporal motifs under specific constraints, such as shared static topology~\citep{sarpe2021oden} or specific input graph structure~\citep{Pu2023SamplingButterfly}. To the best of our knowledge, the only sampling-based algorithm designed for a \emph{streaming setting}, is \EWS by \cite{wang2022efficient}. \EWS uses edge sampling to obtain a relative $\varepsilon$-approximation for individual triangle counts. However, \EWS requires substantial computational resources, including memory---hence, \EWS does not scale on large temporal graphs as demonstrated in \Cref{sec:experiments}.

\spara{Algorithms with predictions in static graphs.} 
The \awp framework introduced by~\citet{mitzenmacher2022algorithms} enhances classical combinatorial algorithms with predictors. Predictors can be obtained, for example, from machine learning models trained on historical data. Classical combinatorial algorithms then benefit from predictions by improving their efficiency, e.g., runtime or memory usage, and retaining their worst-case complexity.  This framework has been applied to clustering~\citep{Jiang2021Ofacility,Ergun2021kMeans}, graph problems~\citep{Lattanzi2023Bellman,Davies2023Flows}, and data structures~\citep{Ferragina2020Learned,Mitzenmacher2019Filters,Aamand2023Frequency}, as well as graph search~\citep{banerjee2022graph,depavia2024learning}, online graph coloring~\citep{antoniadis2024online}, and more~\citep{azar2022online,chen2022faster,bernardini2022universal,brand2024dynamic,henzinger2023complexity}. Related to our work, \cite{chen2022triangle} developed a predictor-based sampling algorithm to estimate triangle or four-cycle counts in \emph{static} graphs over various streaming models. In addition to being restricted to static graphs, the work of \cite{chen2022triangle} relies on the impractical design of querying a perfect predictor that has access to the number of triangles an edge participates in. Clearly, such assumption is far from practical:  $i$) a perfect predictor can be obtained only by solving the triangle counting problem exactly; and $ii$) it cannot model the complex predictors used in practice. Recently, \cite{boldrin2024fast} improved the work of~\citet{chen2022triangle} and proposed a simple, domain-independent predictor that can be obtained with a single pass over the stream. 
We consider the predictor of~\citet{boldrin2024fast} in \Cref{sec:ablation}, showing that such idea is not suitable for \emph{temporal} triangle counting and for our algorithm~\algName. 
In contrast, the predictor we design in \Cref{sec:mdp} leads to very accurate estimates with remarkably small memory usage, especially on massive datasets, solving~\Cref{prob:ourProb}. 

\section{Conclusion} \label{conclusion}

This work addresses the problem of counting temporal triangles in a stream of temporal edges. We introduced \algName, a sampling algorithm enhanced with a predictor, which provides highly accurate estimates while using minimal computational resources compared to current \sota approaches. To the best of our knowledge, \algName is the first algorithm to estimate temporal triangle counts using predictions. Experimental results show that \algName is significantly faster than \sota exact methods and requires less time and memory than approximate \sota streaming algorithms, often obtaining more accurate estimates. Finally, we show how to efficiently compute a simple and practical predictor and evaluate a setting where predictions are leveraged for online processing. 

Future research includes the development of improved and domain-dependent predictors (e.g., learning specific edge-weights for the classification of important edges to retain over the stream) and extending \algName's approach to other temporal motifs, such as butterflies~\citep{Pu2023SamplingButterfly} or motifs sharing a common structure~\citep{sarpe2021oden}.

\section{Acknowledgments}\label{sec:acks}
This research is funded by the Ministry of University and Research within the Complementary National Plan PNC-I.1 ``Research initiatives for innovative technologies and pathways in the health and welfare sector, D.D. 931 of 06/06/2022, PNC0000002 DARE - Digital Lifelong Prevention CUP: B53C2200644000'' and PRIN Project n. 2022TS4Y3N ``EXPAND: scalable algorithms for EXPloratory Analyses of heterogeneous and dynamic Networked Data''. This research is also funded by the
ERC Advanced Grant REBOUND (834862),  and 
the Wallenberg AI, Autonomous Systems and Software Program (WASP) funded by the Knut and Alice Wallenberg Foundation.

\bibliographystyle{abbrvnat}
\bibliography{bibliography}

\appendix
\section{Notation and symbols}\label{appSec:notation}

	\Cref{tab:notation} summarizes the notation and symbols used.
	
	\begin{table}[t]
	\centering
	\caption{Table of notation and symbols used in our work.}
	\label{tab:notation}
	%\resizebox{\columnwidth}{!}{
		\begin{tabular}{ll}
			\toprule
			\bf{Symbol} & \bf{Definition}\\
			\midrule
			$G = (V, E)$ & Temporal graph $G$ with set of vertices $V$ and set of edges $E$ \\
			$(u,v,t)$ & Temporal edge from node $u$ to node $v$ at time $t$\\
			$\tau$ & Time-ordered temporal graph stream \\
			%$T$ & Generic temporal triangle\\
			$T_i$ & Temporal triangle of type $i$ with $i \in [8]$\\
			$\delta$ & Time-duration of $\delta$-instance of a temporal triangle\\
			$\triType_i$ & Set of $\delta$-instances of triangle $T_i$ in $G$, $i \in [8]$\\
			$\estimate_i$ & Estimate of $|\triType_i|$, $i \in [8]$\\
			$\oracle{\cdot}$ & Predictor used by \algName \\
			$p$ & Sampling probability for \algName \\
			% $\EWSp$ & Edge sampling probability of \EWS algorithm \\
			% $\EWSq$ & Wedge sampling probability of \EWS algorithm \\
			% $\NaiveSp$ & Sampling probability of \NaiveS algorithm \\
			$K$ & Number of edges classified as \emph{heavy} by the predictor\\
			$H$ & Set of heavy edges retained by \algName\\
			$S_L$ & Set of light edges sampled by \algName\\
			%$\bigO(\cdot)$ & Big-O notation\\
			%$\wedgeSet$ & Set of wedges\\
			$\wedgeSet_{H,H}$ & Set of wedges with both edges in $H$\\
			$\wedgeSet_{S_L, S_L}$ & Set of wedges with both edges in $S_L$\\
			$\wedgeSet_{H,S_L}$ & Set of wedges with one edge from $H$ and one from $S_L$\\
			% $Bern(p)$ & Bernoulli random variable of parameter $p$\\
			%$\triType_i^{H,H}$ & Subset of $\triType_i$ where the first two edges are heavy \\
			%\triType_i^{L,L}$ & Subset of $\triType_i$ where the first two edges are light \\
			%$\triType_i^{L,H}$ & Subset of $\triType_i$ where exactly one of the first two edges is heavy \\
			$m_\delta$ & Maximum number of edges occurring in any $\delta$-time in \stream\\
			$\edgeWeight{e}{i}$ & Number of $\delta$-instances in $\triType_i$, $i \in [8]$ containing edge $e\in E$ \\
			$\totEdgeWeight{e}$ & Sum of $\edgeWeight{e}{i}$ over all triangles $i \in [8]$ \\
			$\triangleVec$ & Edges in $E$ ordered by non-increasing values according to $\totEdgeW(\cdot)$ \\
			$\topKPerfect$ &Top-$K$ edges of $\triangleVec$, according to $\totEdgeW(\cdot)$ \\
			$\temporalDeg{u, t_a, t_b}$ & Temporal degree of node $u$ within time window $[t_a, t_b]$\\
			$\edgeTemporalDeg{e}$ & Temporal min-degree weight of an edge $e$\\
			$\topKTemporalDeg{K}$ & Top-$K$ edges ordered by non-increasing values according to $\edgeTemporalDeg{\cdot}$\\
			$\tau^{tr}$ & Training stream\\
			$\tau^{ts}$ & Test stream\\
			$\predThreshold$ & Threshold value learned on $\tau^{tr}$\\
			%\midrule
			\bottomrule
		\end{tabular}
	%}
\end{table}

	\section{Missing proofs}\label{appsec:missingProofs}
	
	We now provide the proofs missing from the main text. First, we prove \Cref{unbiasness_theorem}.
	
	\begin{proof}[Proof of \Cref{unbiasness_theorem}]
		
		Fix $i \in [8]$. We partition the set of triangle $\delta$-instances $\triType_i$ as $ \triType_i^{H,H},  \triType_i^{L,H},$ and $\triType_i^{L,L}$, according to the classification of the first two edges $e_1, e_2$ on the stream forming each triangle $\tri = \langle e_1, e_2, e_3 \rangle \in \triType_i$ by \oracle{\cdot}. That is: $\triType_i^{H, H}$ is the set of all the triangles in $\triType_i$ whose first two edges on the stream are classified \emph{heavy} by the predictor (i.e., $\oracle{e_1}=\oracle{e_2}=1$); $\triType_i^{L,L}$ is the set of all triangles in $\triType_i$ whose first two edges are classified as light by the predictor ((i.e., $\oracle{e_1}=\oracle{e_2}=0$); while $\triType_i^{L,H}$ is the set of all triangles in $\triType_i$  with \emph{exactly} one of the first two edges on the stream flagged heavy, (\emph{independently} of the order over the stream).
		Note that since $ \triType_i^{H,H},  \triType_i^{L,H},$ and $\triType_i^{L,L}$ is a partition of $\triType_i$, it holds that $|\triType_i| = |\triType_i^{H,H}| + |\triType_i^{L,H}| + |\triType_i^{L,L}|.$ Recall, that the estimate $\estimate_i$ returned by \algName (\Cref{code:output} of \Cref{alg:MainAlgorithm}) is $\estimate_i = \frac{\estimate_{i,0}}{p^2} + \frac{\estimate_{i,1} }{p}+ \estimate_{i,2}$.  Clearly, $\expectation[\estimate_{i,2}] = \sum_{\tri \in \triType_i^{H,H}}{1} = |\triType_i^{H,H}|$ as the first two edges for each $\tri\in \triType_i^{H,H}$ are deterministically stored in memory by~\algName. Moreover
		\[
		\mathbb{E}[\estimate_{i,1}] = \sum_{\tri \in \triType_i^{L,H}}{\mathbb{E}[X_{\tri}]} = \sum_{\tri \in \triType_i^{L,H}}{p} = p|\triType_i^{L,H}|,
		\]
		using the linearity of expectation and $0-1$ random variables $X_\tri$ for every triangle $\tri \in \triType_i^{L,H}$, with $X_\tri=1$ if the edge $e$ such that $\oracle{e}=0$ (among the first two edges on the stream forming $\tri$) is sampled, which occurs with probability $p$. Analogously, we have
		$\mathbb{E}[\estimate_{i,0}] = \sum_{\tri \in \triType_i^{L,L}}{\mathbb{E}[X_{\tri}]} = p^2 |\triType_i^{L,L}|,$
		where $X_\tri=1$ if both the first two edges of $\tri$ on the stream are sampled, which occurs with probability $p^2$. Combining the above results we get
		$\mathbb{E}[\estimate_i] = 	 p^{-2}\mathbb{E} \left[ \estimate_{i,0}\right] + p^{-1}\mathbb{E}\left[\estimate_{i,1}\right]  +\mathbb{E}[{\estimate_{i,2}}]  =|\triType_i^{L,L}|+ |\triType_i^{L,H}| + |\triType_i^{H,H}|  = |\triType_i|$ concluding the proof.	
	\end{proof}
	
	We now prove the following result, that is used in the proof of \Cref{theo:noisyGuarantees}.
	
	\begin{lemma}\label{lemma:comb}
		For any $\alpha\ge 1, \alpha\in\mathbb{N}$ it holds that,
		\begin{equation}\label{eq:combinatorialRel}
			\alpha \sum_{g=1}^{\alpha+1}  \left[{{\alpha+1}\choose {g-1}} (g-1)! (2\alpha+1-g)!\right] + \alpha!(\alpha+1)! = (2\alpha+1)!
		\end{equation}
		where $0!=1$, and ${n\choose 0} = 1$.
	\end{lemma}
	
	\begin{proof} 
		The proof is by a combinatorial argument. Fix a vector $x = (1,2,\dots,2\alpha+1)$, then the right-hand side of \Cref{eq:combinatorialRel} is just the number of permutations of such a vector, i.e., all the vectors $\pi \in \Pi(x)$. The first term on the left hand side of \Cref{eq:combinatorialRel} can be written as the number of permutations $\pi \in \Pi(x)$ such that the element in position $g\in\{x_1,\dots,x_{\alpha+1}\}$ from $x$ is the minimum element in $\pi$ over the positions $\alpha+2,\dots,2\alpha+1$ in $\pi$, i.e., $g = \min_{i=\alpha+2,\dots,2\alpha+1} \{\pi_i\}$. To count such number of permutations $\pi \in \Pi(x)$, first fix $g\in\{x_1,\dots,x_{\alpha+1}\}$, for such element to hold $g = \min_{i=\alpha+2,\dots,2\alpha+1} \{\pi_i\}$ we construct the permutations as follows i) choose the position of the element $g$ ($\alpha$ possible choices), ii) fix all the $g-1$ elements smaller than $g$ in the sub-sequence $\pi_1,\dots,\pi_{\alpha+1}$ (${{\alpha+1}\choose {g-1}} (g-1)!$ possible choices) and iii) permute all the remaining elements greater than $g$ in the remaining positions ($(2\alpha+1-g)!$ possible permutations). Clearly summing over all such $g\in\{x_1,\dots,x_{\alpha+1}\}$ we get the first term on the left-hand side in \Cref{eq:combinatorialRel}, the remaining term $\alpha!(\alpha+1)!$ corresponds to all the remaining permutations $\pi\in \Pi(x)$, where no element from the subsequence $x_1,\dots,x_{\alpha+1}$ occurs in a permutation $\pi\in \Pi(x)$ over positions $\pi_{\alpha+2},\dots,\pi_{2\alpha+1}$, concluding the proof.
	\end{proof}
	
	\begin{proof}[Proof of~\Cref{theorem:varianceNoNoise}]
		The proof is inspired by Theorem~1.2 from~\citep{chen2022triangle}. However, in our case, the proof significantly differs, as our predictor is a \emph{ranking predictor}. That is, a ranking predictor for an edge $e\in E$ parameterized by $K>1$ classifies if the edge $e\in E$ is in the first $K$ positions in $\triangleVec$ (i.e., $e$ is a heavy edge). An additional issue is that the total weight $\totEdgeW(\cdot)$ of an edge is defined over \emph{all} triangle types, i.e., $\totEdgeW(e) = \sum_{i=1}^8 \edgeWeight{e}{i}$. Given a parameter $K$, let $\rho_K = \totEdgeW(e_{\prec_{K+1}})$ correspond to the weight of the $K+1$-th edge in $\triangleVec$. Then, an edge $e\in E$ on the stream is classified by a ranking predictor as follows
		\[
		\oracle{e}_K=
		\begin{cases}
			1 & \text{if } \triangleVec(e,\prec)\le K, \text{i.e., } W(e) > \rho_K, \\
			0 & \text{otherwise.} 
		\end{cases}
		\]
		Let us fix a triangle type $\triType_i$ for $i \in [8]$, and let $\triType_{i,0}$ be the triangles of type $i\in [8]$ for which \emph{both} the first two edges appearing over the stream (and forming the triangle) are classified as \emph{light} by the predictor (i.e., $\oracle{e}_K = 0$). Recall also that for an edge $e\in E$ classified as light, it holds that $\totEdgeW(e)\le \rho_K$. Let $\chi_{j,i}$ be an indicator random variable denoting that the first two edges of triangle $\tri_{j} \in \triType_{i,0}$ are added to $S_L$ in~\Cref{code:light_edge_sampling} of \algName. For ease of notation as we work with a fixed $i \in [8]$, we simply write $\chi_j$, when clear from the context. Clearly it holds that $\estimate_{i,0} = \sum_j \chi_{j,i}$, and $\expectation[\chi_j] = p^2$, so $\Var[\chi_j] = p^2-p^4 = p^2(1-p^2)$. 
		Let us fix $j_1\neq j_2$ ranging over the triangles in $\triType_{i,0}$ and compute $\Cov(\chi_{j_1}, \chi_{j_2})$, then
		\[
		\Cov(\chi_{j_1}, \chi_{j_2}) = \expectation[(\chi_{j_1} - p^2)(\chi_{j_2}-p^2)] = \expectation[\chi_{j_1}\chi_{j_2}] - p^4.
		\]
		Therefore,
		\begin{align}
			\Var[\estimate_{i,0}] &= \sum_j \Var[\chi_j] +  \sum_{j_1} \sum_{j_2 \neq  j_1} \Cov(\chi_{j_1}, \chi_{j_2}) \nonumber\\ 
			&= \sum_j p^2(1-p^2)+  \sum_{\substack{\tri_{j_1}, \tri_{j_2} \in \triType_{i,1} \\ j_1 \ne j_2}}\Cov(\chi_{j_1}, \chi_{j_2})\nonumber\\
			&= p^2(1-p^2) |\triType_{i,0}| +  \sum_{\substack{\tri_{j_1}, \tri_{j_2} \in \triType_{i,0} \\ j_1 \ne j_2}}\Cov(\chi_{j_1}, \chi_{j_2}). \label{eq:varianceDefn}
		\end{align}
		We will now bound the covariance term $\Cov(\chi_{j_1}, \chi_{j_2})$ for $ j_1\neq j_2$. First, consider two temporal triangles $\tri_{j_1}=\langle e_1^{j_1},e_2^{j_1},e_3^{j_1}\rangle,$ $ \tri_{j_2}=\langle e_1^{j_2},e_2^{j_2},e_3^{j_2}\rangle \in \triType_{i,0}$ for which it holds $|\{e_1^{j_1},e_2^{j_1}, e_1^{j_2},e_2^{j_2}\}| =3$, that is, the two triangles share \emph{exactly one} edge among their first two edges on the stream. We will denote such case with the following notation: ``$|\tri^{12}_{j_1} \cap \tri^{12}_{j_2}| = 1$''. Clearly, the shared edge has to be classified as light, and for such triangles, it holds:
		\[
		\expectation[\chi_{j_1}\chi_{j_2}] = \Prob[\chi_{j_1}=1 | \chi_{j_2} = 1] \Prob[\chi_{j_2}=1] = p^3.
		\] 
		As $ \Prob[\chi_{j_2}=1] = p^2$, and for the event $\chi_{j_1}=1$ to hold, the edge not shared among the two triangles (belonging to $\tri_{j_1}$) needs to be added to $S_L$, which occurs with probability $p$. Differently from static triangles, two distinct temporal triangle $\delta$-instances can share up to \emph{two} temporal edges (i.e., such instances differ in the timestamp on the third remaining edge); therefore, we also need to account for such a case. Let ``$|\tri^{12}_{j_1} \cap \tri^{12}_{j_2}| = 2$'' be the case that two distinct temporal triangle $\delta$-instances $\tri_{j_1},\tri_{j_2} \in \triType_{i,0}$ share both their first two light edges on the stream, then:
		\[
		\expectation[\chi_{j_1}\chi_{j_2}] = \Prob[\chi_{j_1}=1 | \chi_{j_2} = 1] \Prob[\chi_{j_2}=1] = p^2.
		\] 
		Again, it holds $ \Prob[\chi_{j_2}=1] = p^2$, but since the instances share the two first light edges, then $\Prob[\chi_{j_1}=1 | \chi_{j_2} = 1] =1$. 
		Combining everything together, we have that:
		\[
		\Var[\estimate_{i,0}] = p^2(1-p^2) |\triType_{i,0}| +  \sum_{\substack{\tri_{j_1}, \tri_{j_2} \in \triType_{i,0} \\ |\tri^{12}_{j_1} \cap \tri^{12}_{j_2}| = 1}}p^3(1-p) +  \sum_{\substack{\tri_{j_1}, \tri_{j_2} \in \triType_{i,0} \\ |\tri^{12}_{j_1} \cap \tri^{12}_{j_2}| = 2}}p^2(1-p^2).
		\]
		Let us bound the contributions of each term individually. First let $|\maxTriangleWeight| = \sum_{i=1}^8 |\triType_i|$, i.e., the sum of all triangle counts of each type, and let $E(k) = \{e\in E: \totEdgeW(e)=k\}$ be the set of edges that participate in exactly $k$ temporal triangle $\delta$-instances (across all types), then note that $\sum_{k\ge 1} k |E(k)| \le 3 |\maxTriangleWeight|$ as each instance is counted three times (one for each one of its edge). Then,
		\begin{align}
			\sum_{\substack{\tri_{j_1}, \tri_{j_2} \in \triType_{i,0} \\ |\tri^{12}_{j_1} \cap \tri^{12}_{j_2}| = 1}} 1 &\le \sum_{\substack{e:\\ \totEdgeW(e)\le\rho_K}} \edgeWeight{e}{i}^2 \le \sum_{\substack{e:\\ \totEdgeW(e)\le\rho_K}} \totEdgeW(e)^2 \le \nonumber \\
			& \le \rho_K \sum_{k=1}^{\rho_K} k |E(k)| \le 3 \rho_K  |\maxTriangleWeight|, \label{eq:BoundCase1OneShared}
		\end{align}
		where the first inequality comes from summing over all possible light edges, the second inequality comes from the definition of $W(e)$, the third inequality comes from summing over the edge-weights for light edges and that $W(e)\le \rho_K$ for a light edge, and the last inequality comes from $\sum_{k\ge 1} k |E(k)| \le 3 |\maxTriangleWeight|$ as discussed.
		Next,
		\begin{align*}
			\sum_{\substack{\tri_{j_1}, \tri_{j_2} \in \triType_{i,0} \\ |\tri^{12}_{j_1} \cap \tri^{12}_{j_2}| = 2}} 1 
			%&\le \sum_{e: W(e) \le \rho_K} \sum_{e': W(e') \le \rho_K}\min\{W(e),W(e')\}^2 \le \\  
			&\le \sum_{\substack{e: \\ W(e) \le \rho_K}} \sum_{\substack{e': W(e') \le \rho_K \\|t(e')-t(e)| < \delta }}  W(e)^2 \le 6\rho_K |\maxTriangleWeight| m_\delta,
		\end{align*}
		where the first equation comes from bounding the number of possible triangle $\delta$-instances sharing two edges considering that for two temporal triangles to share such edges their timings cannot be far more than $\delta$, and from an analogous argument to \Cref{eq:BoundCase1OneShared} where we let $m_\delta$ be the maximum number of edges in a subgraph spanning an interval of $\delta$. Combining everything together, we have:
		\begin{equation}\label{eq:BoundVarEll1}
			\Var[\estimate_{i,0}] \le((1-p^2)(1+6\rho_K m_\delta) + 3p \rho_K (1-p)) p^2 |\maxTriangleWeight|.
		\end{equation}
		Now let $\triType_{i,1}$ be the set of triangles for which only one of the first two edges arriving on the stream and forming the triangle is classified as light by the predictor (the other one is classified heavy). Let $\xi_{j}$ be an indicator random variable denoting if \emph{only one} of the first two edges of triangle $\tri_{j}\in \triType_{i,1}$ is added to $S_L$ by \algName over the stream. Clearly, $\expectation[\xi_j] = p$, as there is no randomness for the edge classified as heavy. Now let $\estimate_{i,1}$ be the sum over such random variables; it clearly holds $\expectation[\estimate_{i,1}] = p |\triType_{i,1}|$. Furthermore, we now bound the covariance term as for \Cref{eq:varianceDefn}. It clearly holds that $\Cov(\xi_{j_1}, \xi_{j_2}) \le p$ for triangles $\tri_{j_1}\tri_{j_2}\in \triType_{i,1}$ sharing exactly one light edge among their first two edges on the stream.
		We have the following:
		\begin{equation}\label{eq:BoundVarEll2}
			\Var[\estimate_{i,1}] \le p |\triType_{i,1}| + \sum_{\substack{\tri_{j_1}, \tri_{j_2} \in \triType_{i,1}\\ \tri^{12}_{j_1} \cap \tri^{12}_{j_2} = e, \oracle{e}_K =0}} p \le (1+3\rho_K)p|\maxTriangleWeight|
		\end{equation}
		Now, let $\estimate_{i,2}=|\triType_{i,2}|$ be the number of triangles for which both two first edges on the stream are classified as heavy, for which clearly there is no randomness involved. Let $\estimate_i=p^{-2}\estimate_{i,0}+p^{-1}\estimate_{i,1}+\estimate_{i,2}$ be the output of \algName. Then, by combining \Cref{eq:BoundVarEll1} and \Cref{eq:BoundVarEll2} we have:
		\begin{align}
			\Var[\estimate_i] = \Var[p^{-2}\estimate_{i,0}+p^{-1}\estimate_{i,1} &+ \estimate_{i,2}] \le \nonumber \\ 2(( (1-p^2)(1+6\rho_K m_\delta) &+  3p \rho_K (1-p)) \tfrac{|\maxTriangleWeight|}{p^2} + (1+3\rho_K) \tfrac{|\maxTriangleWeight|}{p}) \nonumber \\ & \le 4(p^{-2} + 3p^{-1}\rho_K (p^{-1}m_\delta +1) )|\maxTriangleWeight| \nonumber\\ 
			&\le4 p^{-2} (1 + 6\rho_K m_\delta )|\maxTriangleWeight| \nonumber\\ &\le 28 p^{-2} \rho_K m_\delta |\maxTriangleWeight| \label{eq:BoundVarCase1}
		\end{align}
		
		where we used $\Var[X+Y]\le 2(\Var[X] + \Var[Y])$ for any two random variables $X,Y$, the facts that $\Var[aX] = a^2 \Var[X]$, $1\le p^{-1}m_\delta$ and $1\le \rho_K m_\delta$. By applying Chebyshev's inequality, we obtain: 
		\[
		\Prob[|\estimate_i - |\triType_i|| \ge \varepsilon|\triType_i|] \le \frac{\Var[\estimate_i]}{\varepsilon^2 |\triType_i|^2} \le \frac{28p^{-2}\rho_K m_\delta |\maxTriangleWeight|}{\varepsilon^2 |\triType_i|^2} \le \frac{C_128p^{-2}\rho_K m_\delta}{\varepsilon^2 |\triType_i|}
		\]
		For the sake of the analysis, we assumed that $\frac{|\maxTriangleWeight|}{|\triType_i|^2} = C_1 \frac{1}{|\triType_i|}$ with $C_1$ being a suitable constant, as this is often the case in practice. We leave as future works identifying better bounds to such quantity, e.g., by leveraging the combinatorial structure of the input temporal graph. Also note that $\rho_K$ in practice is generally $\rho_K = |\triType_i|^{\gamma'}, \gamma' \ll 1$, hence $\sqrt{\rho_K} \in \bigO(|\triType_i|^{\gamma'/2})$. Therefore, there exist $\gamma$ such that  setting $p = C_2 \left(\frac{\sqrt{m_\delta}}{\varepsilon |\triType_i|^{1/2-\gamma}}\right)$ the estimates $\estimate_i \in (1\pm \varepsilon)|\triType_i|$ with probability at least $23/24$. Additionally, since the total space used by the algorithm is bounded by $\bigO(m_\delta p + K)$, as trivially we do not need to store edges that dist more than $\delta$ from the current one observed on the stream, hence by setting $K = o(m_\delta)$ the total space used is $\bigO\left(m_\delta \left(\frac{\sqrt{\rho_K m_\delta}}{\varepsilon \sqrt{|\triType_i|}}\right) + K\right) = \bigO\left(m_\delta^{3/2} \left(\frac{1}{\varepsilon |\triType_i|^{1/2-\gamma}}\right)\right)$. Now let $F_i \doteq \estimate_i \notin (1\pm \varepsilon) |\triType_i|, i \in [8]$, that is, $F_i$ corresponds to the event that the algorithm fails to report a $(1\pm\varepsilon)$-approximation of the count of the $i$-th triangle then $\Prob[F_i] < 1/24 $, hence by a union bound over the eight triangles we have $\Prob\left[\bigcup F_i\right] \le \sum \Prob[F_i] \le 1/3$. Thus, with probability $> 2/3$, the final estimates $\estimate_i$ are a relative $\varepsilon$-approximation to their respective triangle count.
	\end{proof}
	
	\begin{proof}[Proof of \Cref{coroll:varianceNoNoise}]
		By \Cref{eq:BoundVarCase1} it immediately follows that $\Var[c_i] \le C p^{-2} \rho_K m_\delta |\triType_i|$ setting $C\ge28C_1$ for $C_1 |\triType_i| = |\hat{\triType}|$, yielding the first part of the proof. Additionally by running~\algName without a predictor it is easy to see that $c_i=p^{-2} c_{i,0}$, then according to the proof of \Cref{theorem:varianceNoNoise} when no predictor is used ($\rho_K = |\hat{\triType}|$) it holds that $\Var[c_{i,0}] \le ((1-p^2)(1+6m_\delta) + 3p(1-p)) p^2 |\hat{\triType}|^2$ which leads to $\Var[c_i] \le C_2 p^{-2} m_\delta |\hat{\triType}|^2$ for a suitable constant $C_2$ and setting $C' \ge C_1^2 C_2$ concludes the proof. 
	\end{proof}
	
	\begin{definition}\label{def:alfaBetaBlock}
		Given $\beta\in [1,\numEdgesTot]$ and $\alpha\in[1,\min\{\beta-1,m-\beta\}]$, we say that $\Pi(\{1,\dots,m\})_{(\alpha,\beta)}$ is an \emph{$(\alpha,\beta)$-block permutation} of $\{1,\dots,m\}$ if it holds:
		\begin{align*}
			\Pi(\{1,\dots,m\})_{(\alpha,\beta)} &= \{\pi_1 \oplus \pi_2 \oplus \pi_3 : \pi_1 \in \Pi(\{1,\dots,\beta-\alpha-1\}), \\ \pi_2 &\in \Pi(\{\beta-\alpha, \beta+\alpha\}), \pi_3 \in \Pi(\{\beta+\alpha+1,m\}) \}
		\end{align*}
		where $\oplus$ denotes vector concatenation (e.g., $(a_1,a_2)\oplus (b_1,b_2) = (a_1,a_2,b_1,b_2)$). 
	\end{definition}		
	An $(\alpha,\beta)$-block permutation represents the set of permutations of the set $\{1,\dots,\numEdgesTot\}$ where three blocks of elements are fixed, i.e., the blocks $(1,\ldots,\beta-\alpha-1 \mathbin\Vert \beta - \alpha, \ldots, \beta+\alpha \mathbin\Vert \beta+\alpha+1, \ldots, \numEdgesTot)$, so elements from a block can only be permuted inside the same block.
	
	\begin{proof}[Proof of~\Cref{theo:noisyGuarantees}]
		%We first start by providing the following definition
		%We are now ready to prove our result.
		First, recall from the definition of noisy predictor (see \Cref{sec:oracle}) 
		that an $\alpha$-noisy ranking predictor draws a permutation $\pi$ uniformly from $\Unif(\Pi{(1,\dots,m)}_{(\alpha,K)})$, and outputs the ranking according to $\triangleVec_\pi$, i.e., the optimal ranking $\triangleVec$ permuted according to $\pi$. Drawing $\pi\sim\Unif(\Pi{(1,\dots,m)}_{(\alpha,K)})$ corresponds to independently sample three vectors $\pi_i \sim \Unif(\Pi(S_i))$, $i=1,2,3$ where $S_1 = \{1,\dots,K-\alpha-1\}, S_2 = \{K-\alpha,\dots,K+\alpha\}$, and $S_3 = \{K+\alpha+1,\dots,m\}$, and obtaining $\pi$ by concatenation, i.e., $\pi= \pi_1\oplus \pi_2 \oplus \pi_3$. To provide a bound on the variance under general value of $\alpha\ge 1$, we fix $i\in[8]$ and we will first bound the term $\Var[\estimate_i | \triangleVec_\pi, \pi\sim\Unif(\Pi(1,\dots,m)_{(K,\alpha)})]$ for which we write $\Var[\estimate_i | \pi]$ for ease of notation. To do so, we recall that we can write $\pi= \pi_1\oplus \pi_2 \oplus \pi_3$, and we can only focus on $\pi_2$ as $\pi_1$ and $\pi_3$ do not affect the variance of~\algName under the $\alpha$-noisy ranking predictor. Let $\Pi{(K-\alpha,\dots,K+\alpha)}$ be the set of permutations from which we draw $\pi_2$; for ease of notation, we will change such index set, by drawing $\pi_2$ from $\Pi{(1,\dots,2\alpha+1)}$, as there is a simple bijection to get the desired $\pi_2$ (i.e., summing $K-\alpha-1$ to each entry of the permuted vector). We also let $\Gamma = \Pi{(1,\dots,2\alpha+1)}$. We now show how to partition the set $\Gamma$ to obtain a bound on the variance of the estimates $\estimate_i, i\in[8]$ computed by~\algName under $\pi_2$ being drawn from each partition.
		Let $\Gamma_{g} = \{\pi \in \Gamma : \min_{j=\alpha+2,\dots,2\alpha+1} \{\pi_{j}\}=g\}, g\in\{1,\dots,\alpha+1\}$, corresponding to the subset of permutations of $\Gamma$ where the minimum element over the elements in $\pi_{\alpha+2},\dots,\pi_{2\alpha+1} $ equals $g\in\{1,\dots, \alpha+1\}$. We define $\Gamma_0\doteq \{\pi \in \Gamma : \min_{j=\alpha+2,\dots,2\alpha+1} \{\pi_{j}\}>\alpha+1\}$, i.e., all the permutations $\pi_2$ not containing any element of the original sequence from $1,\dots,\alpha+1$  in a position $\pi_{\alpha+2},\dots, \pi_{2\alpha+1}$. Then we can write: 
		\[
		\Gamma  = \left(\bigcup_{g=1}^{\alpha+1} \Gamma_g \right) \cup \Gamma_0,
		\]
		that is, the sets $\Gamma_g, g=0,\ldots,\alpha+1$ form a partition of $\Gamma$, and we recall that $\Gamma$ corresponds to the set from which $\pi_2$ is drawn uniformly at random. Additionally, for a fixed $g\in \{1,\ldots,\alpha+1\}$ when $\pi_2\sim \Unif(\Gamma_g)$, then by a similar analysis to~\Cref{theorem:varianceNoNoise} we obtain $\Var[\estimate_i | \pi_2\sim \Unif(\Gamma_g)] \le g \boundDiff 28 p^{-2} \rho_K m_\delta |\maxTriangleWeight| $ by noting that:
		\[\sum_{\substack{e' \in \triangleVec_{\pi_2}(j) : j > K \\ \pi_2 \sim \Gamma_g}} \totEdgeW(e')^2 \le
		\rho_K \boundDiff g \sum_{k\ge 1} k |E(k)| \le 3 \rho_K  \boundDiff g |\maxTriangleWeight|,
		\]  
		that is if $\pi_2\sim \Gamma_g$ then the maximum weight $W(e')$ over the entries $e' \in \triangleVec_{\pi_2}(j), j> K$ is bounded by $g\boundDiff$. While for $\pi_2\sim \Unif(\Gamma_0)$ it holds $ \Var[\estimate_i | \pi_2\sim \Unif(\Gamma_0)]  \le 28 p^{-2} \rho_K m_\delta |\maxTriangleWeight| $. Let
		\begin{align}
			&\expectation_{\pi\sim \Gamma}[\Var[\estimate | \pi]] = \sum_{g=0}^{\alpha+1} \Var[\estimate| \pi \sim \Gamma_g] \Prob[\pi \in \Gamma_{g}]  \le \nonumber\\
			& \le 28 p^{-2} \rho_K m_\delta |\maxTriangleWeight|  \boundDiff_\alpha \sum_{g=0}^{\alpha+1} \Prob[\pi \in \Gamma_{g}]\label{eq:boundCondVar}.
		\end{align}
		Yielding therefore that $\expectation_{\pi\sim \Gamma}[\Var[\estimate | \pi]] \le 28 p^{-2} \rho_K m_\delta |\maxTriangleWeight|  \boundDiff_\alpha $ where such bound comes by noting that $\Gamma_g, g\ge0$ is a partition of $\Gamma$ or alternatively, using \Cref{lemma:comb} and recalling that $\boundDiff_\alpha = \boundDiff (\alpha+1)$. Now, using the law of total variance, we have the following,
		\[
		\Var[c_i] = \expectation_{\pi\sim \Gamma}[\Var[\estimate_i | \pi]]  + \Var_{\pi\in \Gamma}[\expectation[\estimate_i|\pi]] \le 28 p^{-2} \rho_K m_\delta |\maxTriangleWeight|  \boundDiff_\alpha
		\]
		where the last inequality follows from the fact that $\expectation[\estimate_i]$ equals $\expectation_{\pi\in \Gamma}[\expectation[\estimate_i | \pi]] = |\triType_i|$ combined with the bound obtained in \Cref{eq:boundCondVar}. We conclude the proof by the same argument as in \Cref{theorem:varianceNoNoise} and the choice of $p$ as in the statement.
	\end{proof}

	\section{Time complexity}
	\label{appendix:time_complexity}
	
	We now analyze the time complexity of~\algName presented in \Cref{sec:oracle}. Let us recall that $m_{\delta}$ denotes the maximum number of edges of the temporal graph within a time-duration $\delta$ and that we denote with $d(u,t_a, t_b)$ the \emph{temporal degree} of node $u$ within the time window $[t_a,t_b]$. 
	Clearly, given a node $u\in V$ then $d(u,t - \delta, t)\le m_\delta$.\footnote{Sharper bounds can be obtained by using different properties, e.g., the maximum degree in a time-duration $\delta$, we use such bound to obtain a simple bound on the time complexity.} 
	The time required by \algName to process each temporal edge $e\in \tau$ over the stream (\Cref{code:time_pruning} - \Cref{code:light_edge_sampling}) is dominated by the function \wedgeRoutine. 
	In fact, the time complexity of \texttt{CleanUp} (\Cref{code:time_pruning}) is at most $\bigO(m_\delta)$, and the combined time complexity of the calls to $\updateRoutine$ (\Cref{code:update_estimate_1} - \Cref{code:update_estimate_3}) is bounded by the time complexity to execute \wedgeRoutine as each wedge processed by $\updateRoutine$ is processed in $\bigO(1)$ time. It is easy to see that for a given $e=(u,v,t)$ over the stream $\tau$ the \wedgeRoutine procedure collects at most $d(z, t - \delta, t)^2 \le m_\delta^2$ wedges for $z=\arg\max\{d(u,t -\delta, t),d(v,t -\delta, t)\}$, where each wedge is collected in constant time. Additionally, each call to the predictor $\oracle{\cdot}$ (\Cref{code:heavy_classification}) requires constant time. We provide further details on the \wedgeRoutine subroutine in \Cref{appendix::subroutines} and on our implementation in \Cref{sec:experiments}. Therefore, by noting that when \algName is executed with a sampling probability $p$ and a $K$ ranking predictor $\oracle{\cdot}_K$ (see \Cref{sec:oracle}) for $K=o(m_\delta)$, the sizes of $S_L$ and $H$, at each iteration, are bounded in \emph{expectation} by $p m_{\delta}$ and $|H| \le K$ respectively.\footnote{Recall that, at each time step, the set $H$ stores the edges predicted heavy by the predictor $\oracle{\cdot}_K$, thus the bound $|H| \le K$ always holds.} This yields to the final expected time complexity $\bigO(m (pm_{\delta} + |H|)^2)$.
	
	\section{Subroutines}\label{appendix::subroutines}
	
	In this section, we describe the subroutines used in \Cref{alg:MainAlgorithm}, namely, \wedgeRoutine and \updateRoutine subroutines. We will first start by introducing the \wedgeRoutine procedure.
	
	\begin{algorithm}[t]
		\KwIn{Heavy edge set $H$, light edge set $S_L$, current edge $e=(u,v,t)$ on $\tau$.}
		\KwOut{Sets $\wedgeSet_{H,H}, \wedgeSet_{H,S_L}, \wedgeSet_{S_L, S_L}$ of wedges built from nodes $u$ and $v$.}
		
		$\wedgeSet_{H,H} \leftarrow \varnothing$; $\wedgeSet_{H,S_L} \leftarrow \varnothing$; $\wedgeSet_{S_L,S_L} \leftarrow \varnothing$\;
		$u,v,t \leftarrow e$\;
		$x \leftarrow \argmin{ \temporalDeg{u, t-\delta, t}, \temporalDeg{v, t-\delta, t}}$\; \label{wedgeRoutine::mindeg}
		$y \leftarrow \argmax{ \temporalDeg{u, t -\delta, t}, \temporalDeg{v, t-\delta, t}}$\; \label{wedgeRoutine::maxdeg}
		Let $N(x)$ be the set of nodes connected to $x$ within the time window $[t - \delta, t]$\;
		$\tilde{N}(x) \leftarrow \{(x,z_i,t_j), (z_i,x,t_j) \in H \cup S_L \ | \  z_i \in N(x) \setminus \{y\} \}$\; \label{wedgeRoutine::neighborhood}
		
		\ForEach{$e_x = (u_x, v_x, t_x) \in \tilde{N}(x)$}{
			%Let $e_x$\;
			\If {$u_x = x$} { $\tilde{E}(y) \leftarrow \{(y, v_x, t_j), (v_x, y, t_j) \in H \cup S_L\}$\;} \label{wedgeRoutine::source} 
			\Else{$\tilde{E}(y) \leftarrow \{(y, u_x, t_j), (u_x, y, t_j) \in H \cup S_L\}$\;} \label{wedgeRoutine::target} 
			
			\ForEach{$e_y\in \tilde{E}(y)$} {
				\lIf{$t_x < t_y$\label{wedgeRoutine::check_time}}{$w = \langle e_x, e_y \rangle$} 
				\lElse{ $w = \langle e_y, e_x \rangle$}
				\lIf{$e_x \in H \land e_y \in H$} {$\wedgeSet_{H,H} \leftarrow \wedgeSet_{H,H} \cup \{w\}$} \label{wedgeRoutine::HH}
				\lElseIf{$e_x \in H \lor e_y \in H$} {$\wedgeSet_{H, S_L} \leftarrow \wedgeSet_{H, S_L} \cup \{w\}$} \label{wedgeRoutine::HL}
				\lElse{$\wedgeSet_{S_L,S_L} \leftarrow \wedgeSet_{S_L,S_L} \cup \{w\}$} \label{wedgeRoutine::LL}
			}
		}
		\KwRet{$\wedgeSet_{H,H}, \wedgeSet_{H,S_L}, \wedgeSet_{S_L,S_L}$.}
		\caption{\wedgeRoutine($H, S_L, e$)}\label{wedgeRoutine}
	\end{algorithm}
	
	\spara{\wedgeRoutine}. The pseudocode for \wedgeRoutine is given in ~\Cref{wedgeRoutine}. Given as input the set $H$ of heavy edges, the set $S_L$ of light edges and the current edge $e = (u,v,t)$ on the stream, \wedgeRoutine returns the sets  $\wedgeSet_{H,H}, \wedgeSet_{H,S_L}, \wedgeSet_{S_L, S_L}$ of wedges that form a temporal triangle with $e$, where each wedge is built starting from the neighborhoods of the nodes $u,v$. 
	\Cref{wedgeRoutine::mindeg} - \Cref{wedgeRoutine::maxdeg} identify the node with the smallest temporal degree within the time window $[t - \delta, t]$ among $u$ and $v$.\footnote{We denote with $d(u,t_a, t_b)$ the temporal degree of a node $u \in V$ within the time window $[t_a, t_b]$} 
	We label such node as $x$, and the remaining node as $y$. 
	In \Cref{wedgeRoutine::neighborhood}, we build the set $\tilde{N}(x)$ of edges adjacent to node $x$ (excluding edge $e = (u,v,t)$). 
	Then, for each edge $e_x = (u_x, v_x, t_x)\in \tilde{N}(x)$, if the source node $u_x$ corresponds to the node $x$, we collect the set $\tilde{E}(y)$ as the set of edges of the form $(y, v_x, t_j)$ and $(v_x, y, t_j)$ (\Cref{wedgeRoutine::source}). 
	If instead it holds that $v_x = x$ (\Cref{wedgeRoutine::target}), we collect the set $\tilde{E}(y)$ by gathering edges including $y$ and $u_x$. 
	Once collected the set $\tilde{E}(y)$, we compare, for each pair of edges $e_x \in \tilde{N}(x)$ and  $e_y \in \tilde{E}(y)$, the  timestamps $t_x$ and $t_y$ of $e_x$ and $e_y$ respectively (\Cref{wedgeRoutine::check_time}): if $t_x < t_y$ we build a wedge of the form $w = \langle e_x, e_y \rangle$, otherwise we have $w = \langle e_y, e_x\rangle$. We then classify the wedge $w$ according to the membership of the edges $e_x, e_y$ to the sets $S_L, H$ (\Cref{wedgeRoutine::HH} - \Cref{wedgeRoutine::LL}).
	
	\spara{\updateRoutine}. The pseudocode for \updateRoutine is given in~\Cref{updateRoutine}. Given the counter $\estimate_i$ associated to the temporal triangle $T_i$, a set of wedges $\wedgeSet$ and an edge $e$, \updateRoutine updates the counter $\estimate_i$ by checking, for each wedge in $\wedgeSet$, if it forms a triangle of type $T_i$ with edge $e$. This is done by a \triangleRoutine procedure (\Cref{updateRoutine::triangleRoutine}), which, given a wedge and an edge, checks which type of temporal triangle they form. Since \triangleRoutine is trivial, we omit its pseudo-code. 
	
	\begin{algorithm}[t]
		\KwIn{Counter $\estimate_i$, wedge set $\wedgeSet$, temporal edge $e$.}
		\KwOut{Updated counter $c_i$.}
		\ForEach{$w \in \wedgeSet$} {
			$\estimate_i \leftarrow \estimate_i + \boldsymbol{1}[\triangleRoutine(w,e) = T_i];$ \label{updateRoutine::triangleRoutine}
		}
		\KwRet{$\estimate_i$.}
		\caption{\updateRoutine($\estimate_i, \wedgeSet, e$)}\label{updateRoutine}
	\end{algorithm}

	\section{Parameters and sensitivity to $K$}\label{appendix::parameters}
	
	\begin{table}
	\centering
	%\captionsetup{width=\textwidth}
	\caption{Sampling probability $p$ of \algName, percentage $K / m$ of edges retained as heavy by~\algName's predictor and the sampling  probability \NaiveSp of \NaiveS for each datasets.}
	\label{tab:parameters}
	%\resizebox{0.35\columnwidth}{!}{
		\begin{tabular}{lcccc}
			\toprule
			\bf{Parameters} & \bf{SO} & \bf{BI} & \bf{RE} & \bf{EC}\\
			\midrule
			$p$ &  $0.1$ & $0.01$ & $0.01$ & $0.01$\\
			$K/m$ & $0.01$ & $0.01$ & $0.01$ & $0.01$\\
			\NaiveSp & $0.109$ & $0.0199$ & $0.0199$ & $0.0199$\\
			\bottomrule
		\end{tabular}
	%}
\end{table}

	In this section, we detail the parameters used in our experiments and provide additional results on the influence of the parameter $K$ on \algName.
	
	\spara{Parameters}. The parameters for \algName and \NaiveS are reported in \Cref{tab:parameters}, and are fixed for all values of the time-window $\delta$ considered. 
	We kept the same parameters for all \algName's variants (i.e., \algNameWTemporal, \algNameWStatic and \algNameWHybrid) when performing the ablation study on \algName's practical predictor (\Cref{sec:ablation}). For \Degen~\citep{pashanasangi2021faster}, since such method can also bound the temporal duration of the inter-time events over each $\delta$-instance, we set its parameter to match the definition of~\citep{paranjape2017motifs} as also done by the authors in their work. 
	All exact parallel algorithms (i.e., \Motto and \FastTri) are executed using a single thread to obtain a meaningful comparison with all other methods, which are also executed sequentially. 
	%We did not set any degree threshold for \FastTri since it is considered only when exploiting multiple threads. 
	For \Motto, based on personal communication with the authors, we set its parameter $\omega$ to 4 (as proposed in the authors' implementation guidelines).\footnote{Such parameter has a small impact on \Motto's performances.} 
	%The \EWS algorithm, 
	%when estimating the count of temporal triangles, performs two different samplings: one to sample edges of the input graph and one to sample wedges that are built from the previously sampled edges. 
	The \EWS algorithm requires two different sampling parameters. We set the edge sampling probability \EWSp and the wedge sampling probability \EWSq to 0.01 and 0.1, respectively, for all datasets, as done by Wang et al.~\citep{wang2022efficient} when dealing with large temporal graphs, for which most of datasets coincide with ours (except the EC dataset). 
	The time durations considered for the datasets SO, BI, and RE are $3\,600$s, $86\,400$s, and $259\,200$s. For the EC dataset, we chose $1 \times 10^5 \mu$s, $2 \times 10^5 \mu$s and $3 \times 10^5 \mu$s, due to the different precision of timestamps.
	
\begin{figure}[h]
		\centering
		\includegraphics[width=\columnwidth]{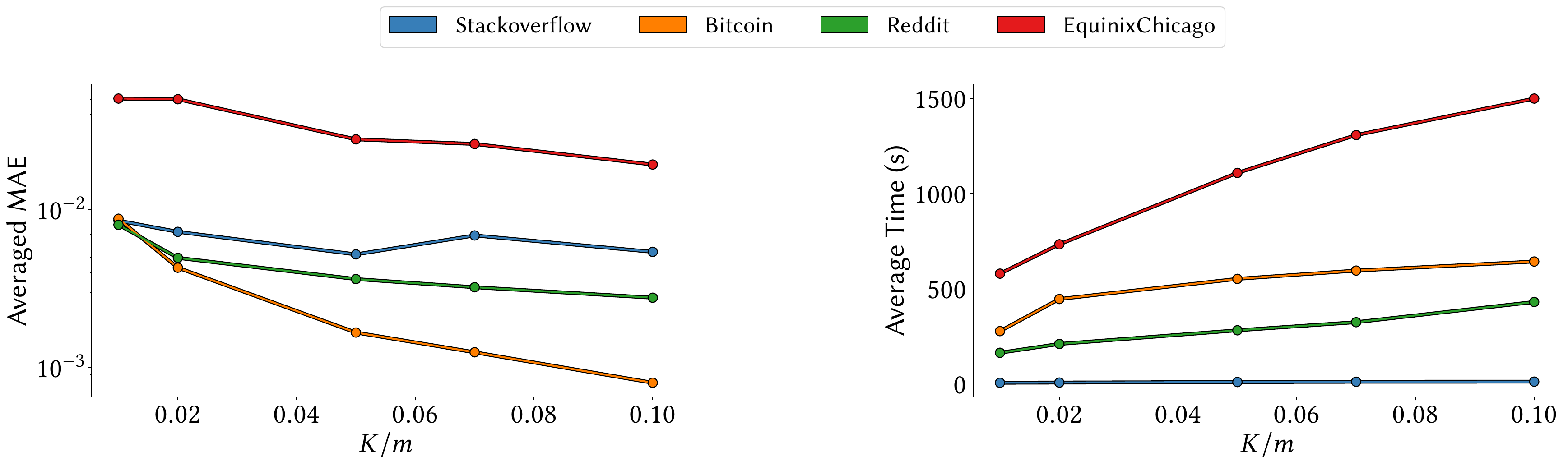} 
		\caption{Left: MAE obtained by \algNameWTemporal averaged over the eight type of triangles (see \Cref{fig:triangles}) and over 10 trials as a function of $K/m$. Y-axis is in log-scale for visualization purposes. Right: corresponding execution times obtained by \algNameWTemporal averaged over 10 trials as a function of $K/m$. Results are for the highest value of $\delta$ on each dataset. }
		\label{fig:error_time_vs_K}
	\end{figure}

	\spara{Sensitivity to $K$}. We performed an experiment to assess the impact of the parameter $K$ on \algNameWTemporal's performances:
	%are affected by the parameter $K$ in terms of 
	as metrics, we consider the quality of the estimates and the running time. On the left of \Cref{fig:error_time_vs_K}, we report the averaged MAE (over 10 trials and over the eight different types of triangles) as a function of $K/m$, that is, the percentage of edges that are classified heavy by the temporal min-degree predictor. On the right, we show the corresponding execution times of \algNameWTemporal (averaged over 10 trials). These results show that higher quality estimates require more runtime by \algNameWTemporal, capturing an important time vs.\ accuracy tradeoff. In time-critical, or online, applications, it is possible to speed-up \algName by decreasing the parameter $K$ at the cost of estimates with a slightly larger variance.
	
	\section{Datasets details}
	\label{appendix::equinix_and_preproc}
	
	In this section, we provide a detailed explanation of how we built the EquinixChicago (EC) dataset and the preprocessing performed for all the datasets considered in our experimental evaluation.
	
	\spara{EquinixChicago dataset}. Let $G=(V,E)$ be the original \emph{bipartite} graph from \citep{sarpe2021presto}. Our goal is to generate a set $E'$ of new edges such that the resulting temporal graph $G'=(V,E \cup E')$ contains $\delta$-instances of temporal triangles. For each node $v \in V$, we uniformly sample a set $X$ of eight nodes that are neighbors of $v$ (i.e., nodes that share an edge with $v$). Then, for each node $x \in X$, we uniformly sample a set $Y$ of eight neighbors of $x$. We then construct the set of wedges $\wedgeSet_{v,x,y}$ by sampling sixteen pairs of temporal edges of the form $\langle (v,x,t_1), (x,y,t_2) \rangle$ (with $v \in V, x \in X, y \in Y$). Finally, for each wedge $\langle (v,x,t_1), (x,y,t_2) \rangle \in \wedgeSet_{v,x,y}$, we generate a temporal edge, which is randomly chosen to be of the form $(v,y,t_3)$ or $(y,v,t_3)$. The time $t_3$ is selected uniformly at random within the time interval $[t_1, t_2]$ (assuming, without loss of generality, that $t_1 < t_2$). Each newly generated edge is added to the set $E'$. After processing each node in $V$, we constructed the final graph as $G' = (V, E \cup E')$. 
	
	\spara{Datasets preprocessing}. For each temporal graph $G =(V,E)$, we perform the following preprocessing steps:
	\begin{enumerate}
		\item We remap each node of the dataset such that each node has a unique integer ID within the interval $[0, n-1]$ (where $n = |V|$, i.e., $n$ is the total number of nodes in the graph).
		\item We remove all self-loops contained in $E$. An edge of the form $(u,u,t)$ (with $u \in V$) is therefore not considered.
		\item We remove all duplicate edges from $E$. %An edge $(u,v,t)$ becomes unique after this processing step.
		\item We sort all the edges according to their timestamps in ascending order. 
	\end{enumerate}
	Note that this is a standard pre-processing performed by all considered baselines. 
	
	\section{Additional results}
	\label{appendix:additional_results}
	
\begin{figure}
		\centering
			\vspace{-1cm}
			\includegraphics[width=0.77\textwidth]{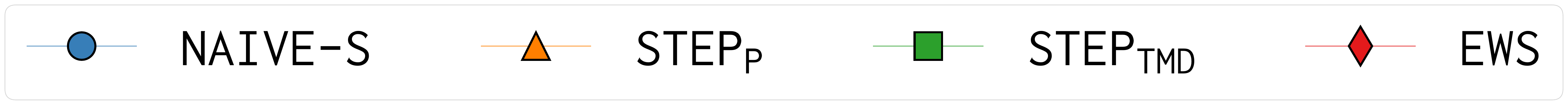}
			\includegraphics[width=0.77\textwidth]{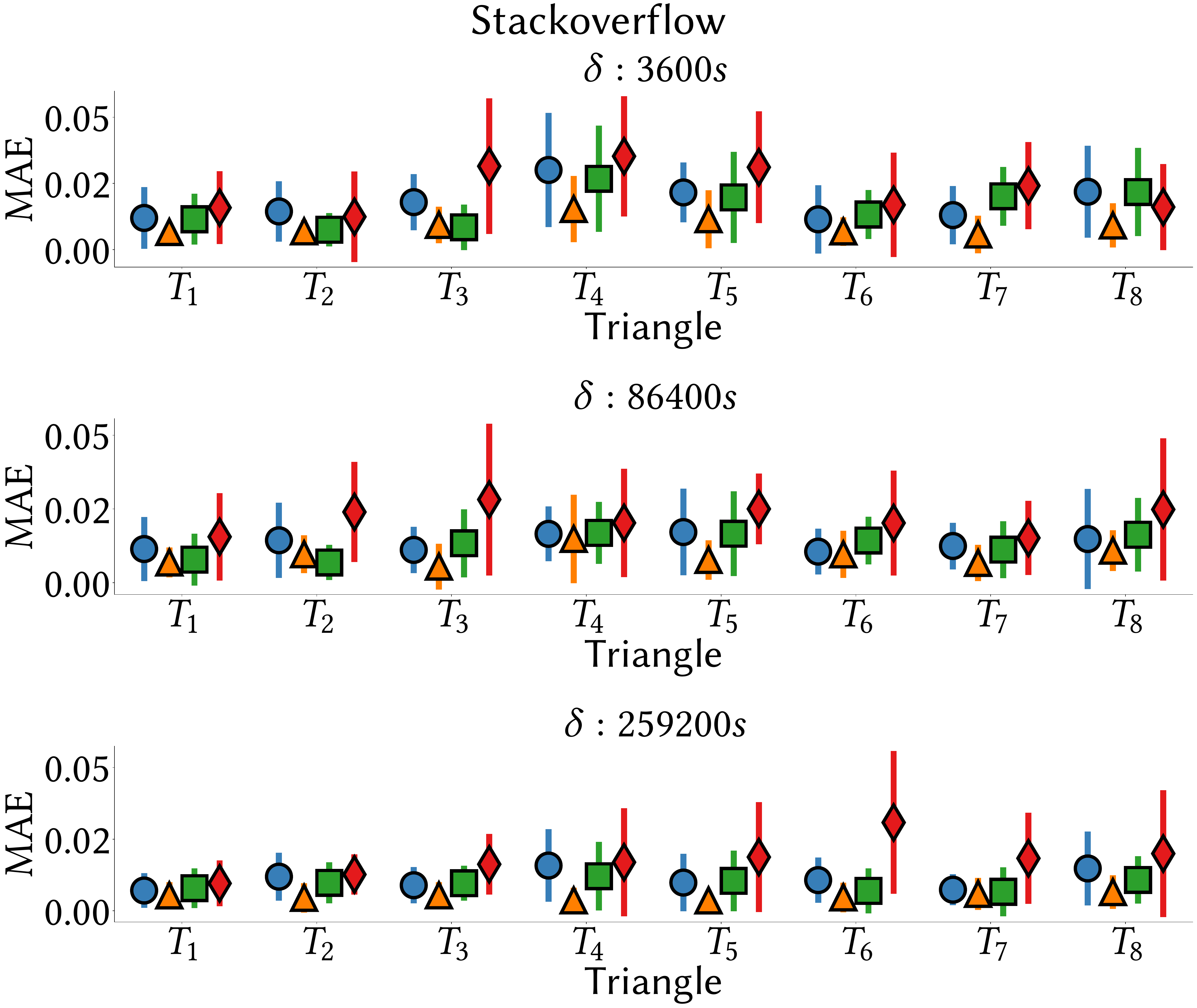}
			\includegraphics[width=0.77\textwidth]{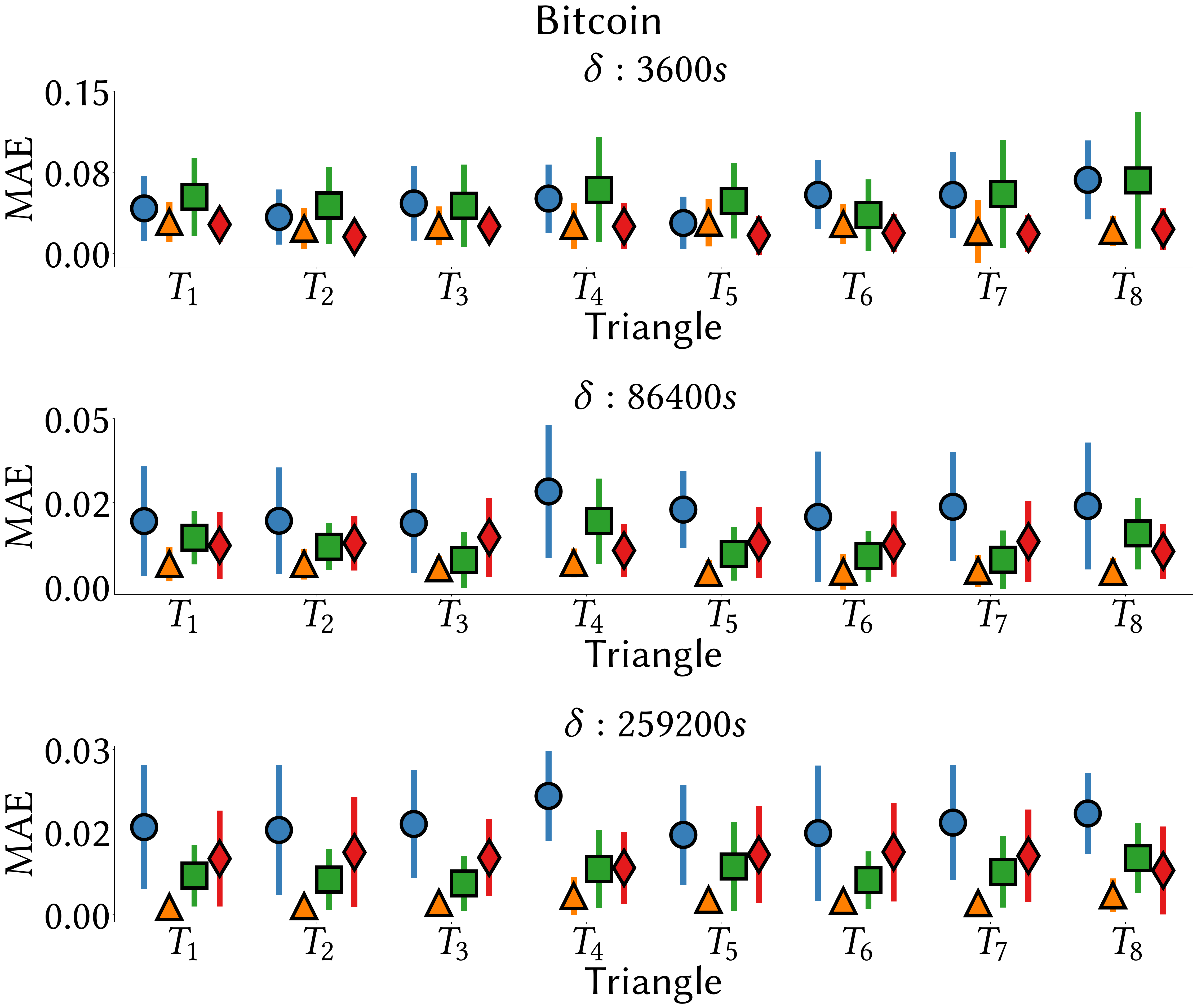}
		\caption{MAE and standard deviation of \algName and baseline algorithms (\NaiveS and \EWS) on SO and BI datasets from \Cref{tab:datasets}, for each value of $\delta$ considered and for each temporal triangle from \Cref{fig:triangles}.}
		\label{fig:error_results_complete_SO_BI}
\end{figure}

\begin{figure}
		\centering
			\vspace{-1cm}
			\includegraphics[width=0.77\textwidth]{res/charts/offline_legend.pdf}
			\includegraphics[width=0.77\textwidth]{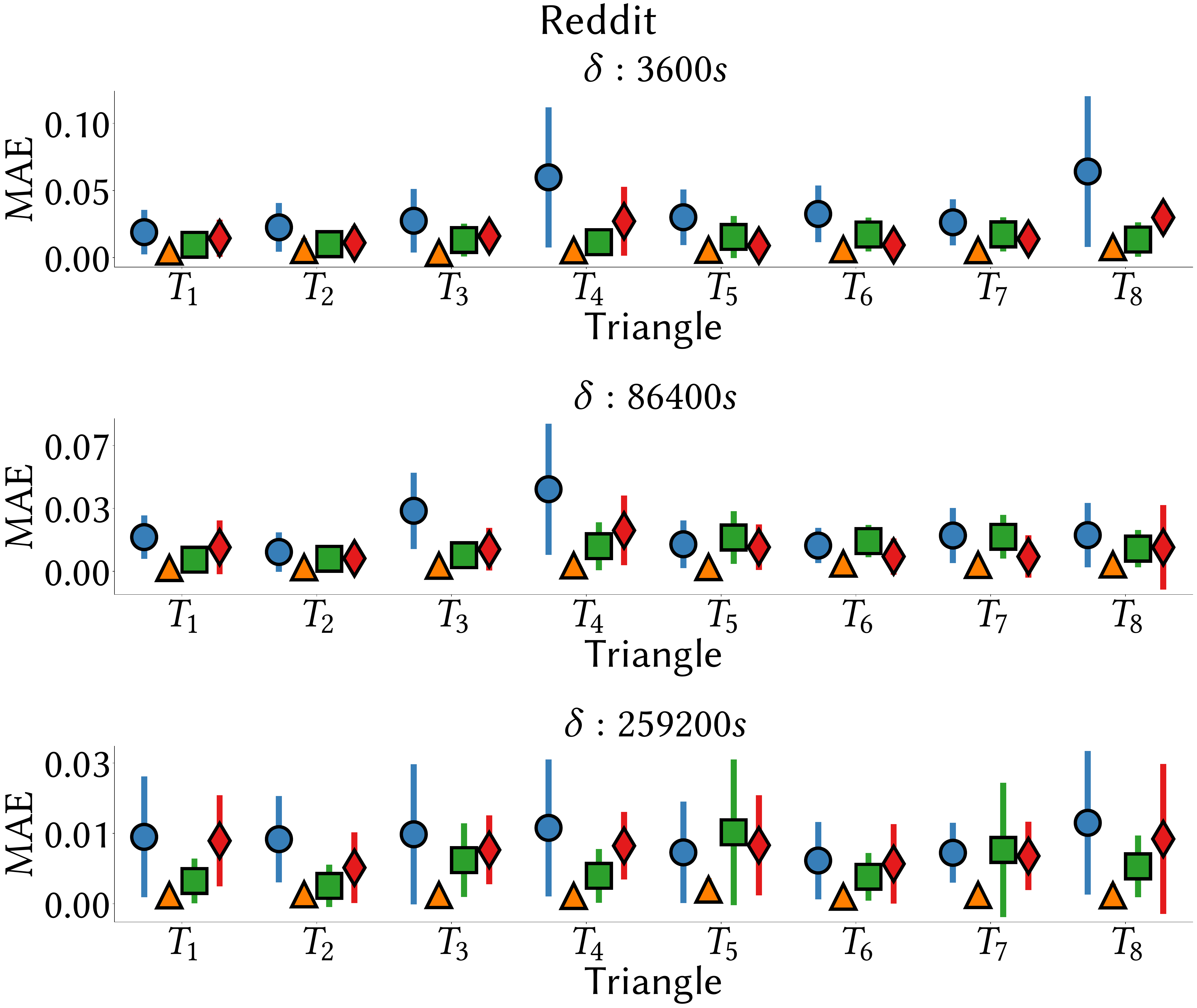}
			\includegraphics[width=0.77\textwidth]{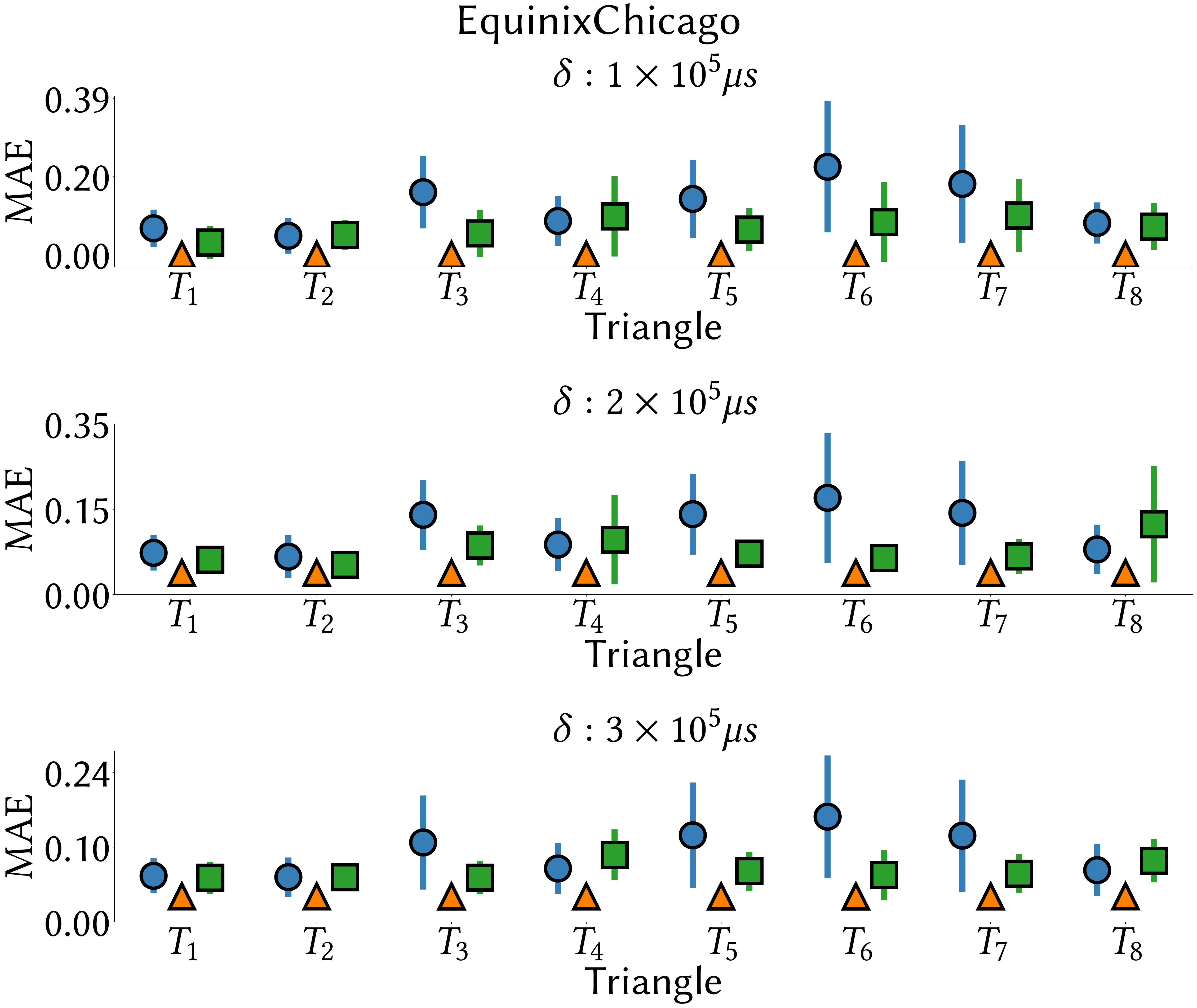}
		\caption{MAE and standard deviation of \algName and baseline algorithms (\NaiveS and \EWS) on RE and EC dataset from \Cref{tab:datasets}, for each value of $\delta$ considered and for each temporal triangle from \Cref{fig:triangles}.}
		\label{fig:error_results_complete_RE_EC}
\end{figure}

\begin{figure}
		\centering
			\vspace{-1cm}
			\includegraphics[width=0.4\textwidth]{res/charts/online_legend.pdf}
			\includegraphics[width=0.77\textwidth]{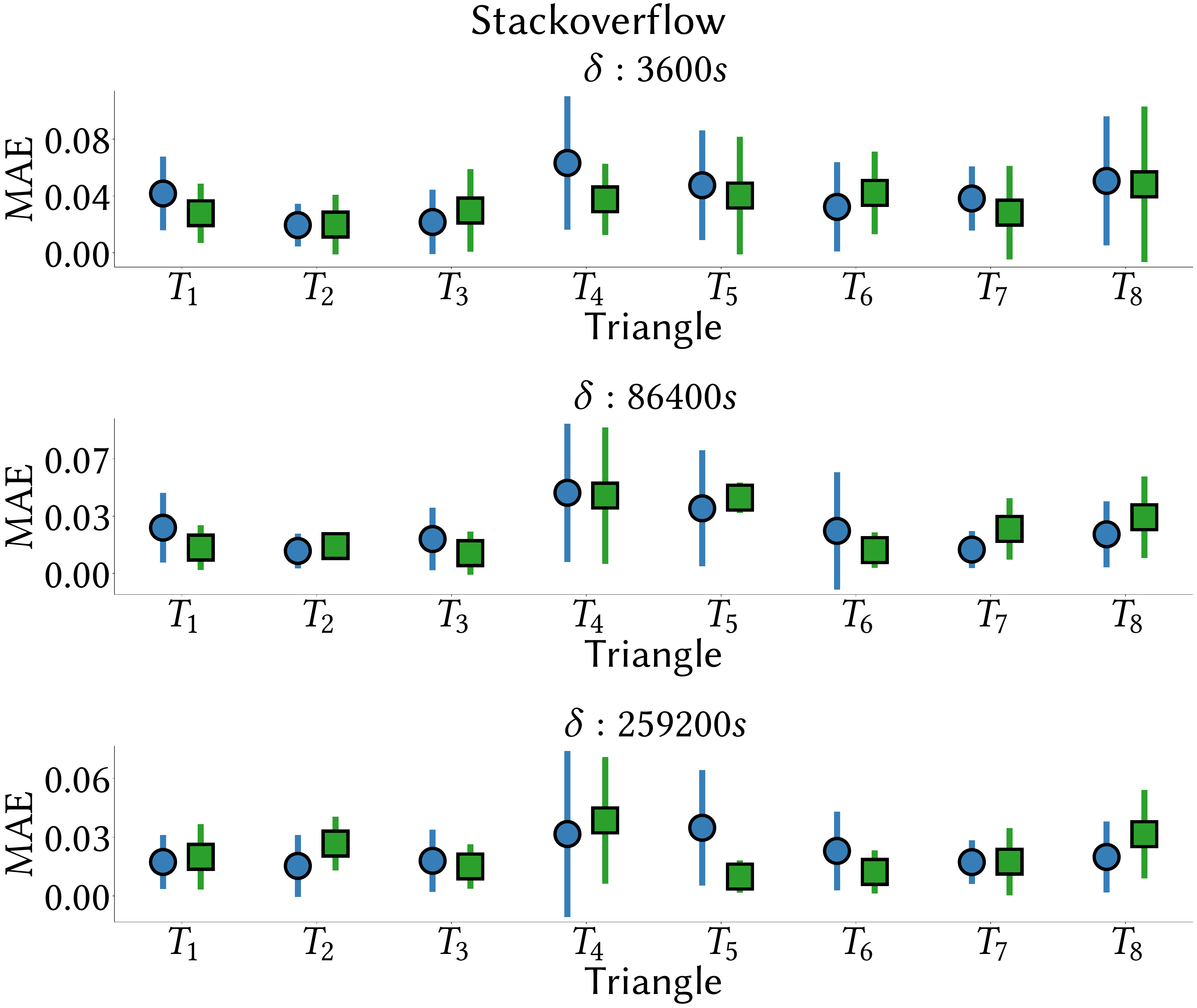}
			\includegraphics[width=0.77\textwidth]{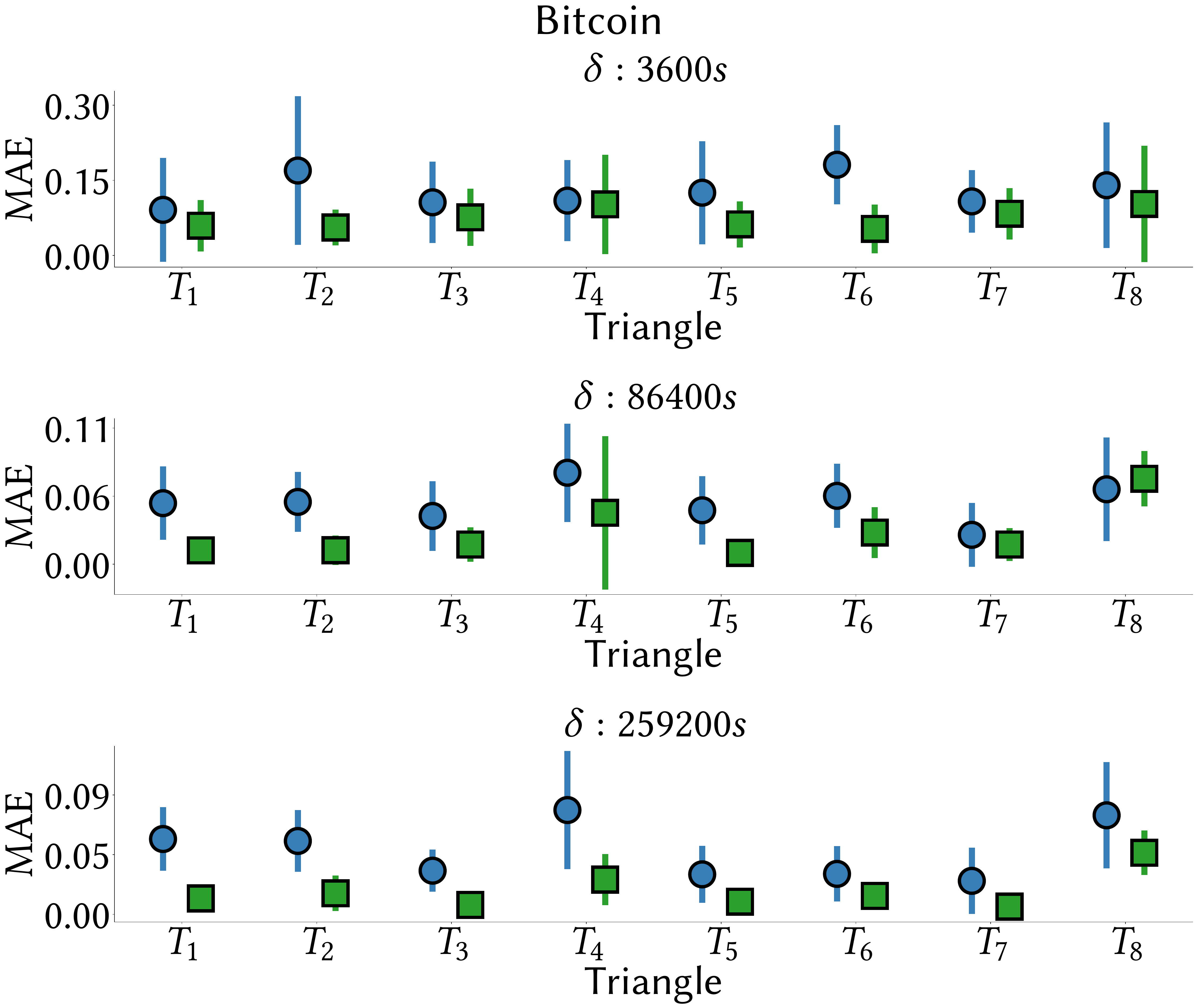}
		\caption{MAE and standard deviation for online estimation on the stream $\tau^{ts}$ by the \NaiveS algorithm and \algNameWTemporal (trained on the historical data from $\tau^{tr}$) on SO and BI datasets, for all values of $\delta$ and for each triangle type (from \Cref{fig:triangles}).}
		\label{fig:error_results_split_stream_complete_SO_BI}
	\end{figure}

\begin{figure}
		\centering
			\vspace{-1cm}
			\includegraphics[width=0.4\textwidth]{res/charts/online_legend.pdf}
			\includegraphics[width=0.77\textwidth]{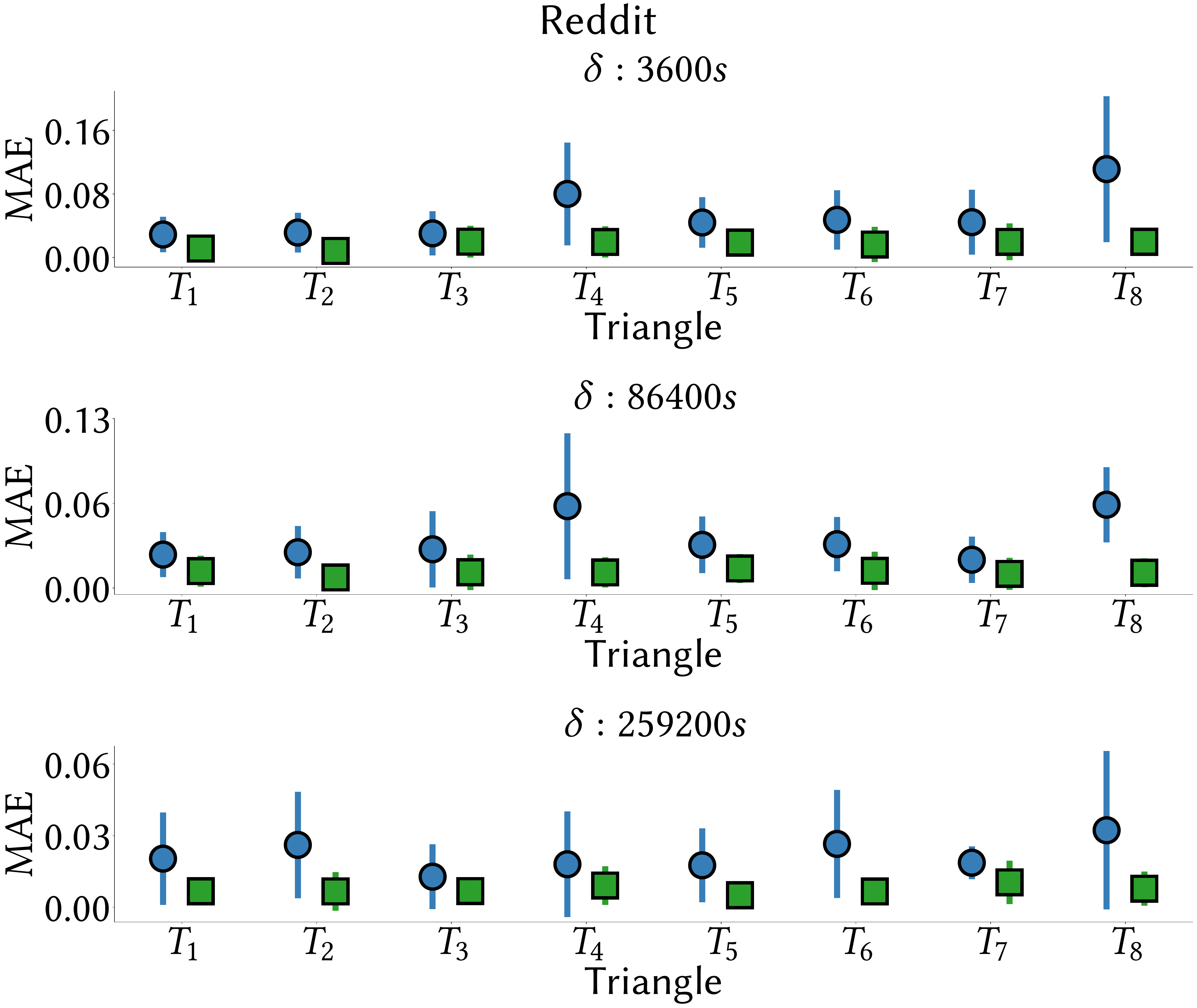}
			\includegraphics[width=0.77\textwidth]{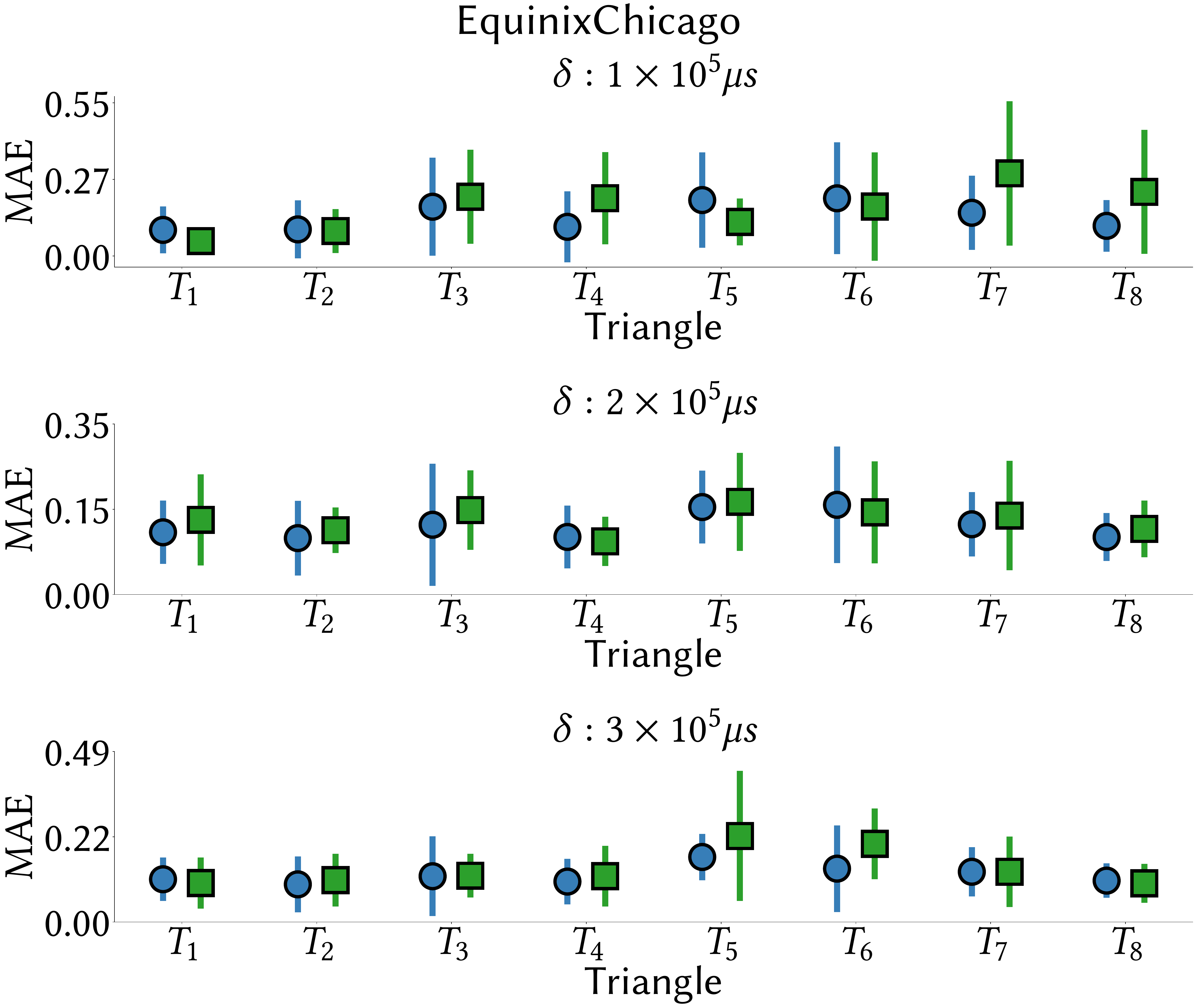}
		\caption{MAE and standard deviation for online estimation on the stream $\tau^{ts}$ by the \NaiveS algorithm and \algNameWTemporal (trained on the historical data from $\tau^{tr}$) on RE and EC datasets, for all values of $\delta$ and for each triangle type (from \Cref{fig:triangles}).}
		\label{fig:error_results_split_stream_complete_RE_EC}
	\end{figure}

	In this section, we complement our experimental evaluation by discussing further results. \Cref{fig:error_results_complete_SO_BI} and \Cref{fig:error_results_complete_RE_EC} show the MAE obtained by all algorithms for all configurations and for all values of the time-duration $\delta$. Table~\ref{tab:time_results} reports the execution time of all algorithms, including the \NaiveS baseline and the \algName algorithm when coupled with a perfect predictor (\algNameWPerfect). Finally, \Cref{fig:error_results_split_stream_complete_SO_BI} and \Cref{fig:error_results_split_stream_complete_RE_EC} show the MAE error obtained by \NaiveS and \algName in the \emph{online} setting described in \Cref{sec:online_est}.
	
	\begin{table}
	\centering
	\caption{Average runtime over ten runs (in seconds). ``\xmark'' denotes out of maximum RAM memory (200 GB). We do not report the standard deviation of exact algorithms as such algorithms were executed only once, given their high runtime on large data. For each configuration, the best runtime is denoted in \textit{italic}, while the second best runtime is highlighted in bold. }
	\label{tab:time_results}
	\resizebox{\textwidth}{!}{
	\begin{tabular}{llccccccc}
	\toprule
	Dataset &
	$\delta$ &
	\NaiveS&
	$\algNameWPerfect$ &
	$\algNameWTemporal$ &
	\EWS &
	\Degen &
	\FastTri & 
	\Motto \\
	\midrule
	\multirow{3}{*}{SO} & $3600s$ & \textit{4.12 $\pm$ 0.02} & $5.29 \pm 0.02$ & \textbf{4.85 $\pm$ 0.02} & $30.37 \pm 0.81$ & $348.41$ & $15.1$ & $174.5$\\
									    & $86400s$ & \textit{5.25 $\pm$ 0.01} & \textbf{6.07 $\pm$ 0.05} & 6.34 $\pm$ 0.1 & $31.96 \pm 0.46$ & $355.15$ & $41.8$ & $254.5$ \\
									    & $259200s$ & \textit{6.62 $\pm$ 0.02} & $7.46 \pm 0.03$ & \textbf{7.45 $\pm$ 0.14} & $35.59 \pm 0.89$ & $356.34$ & $76.3$ & $378.9$ \\
	\midrule
	\multirow{3}{*}{BI} & $3600s$ &  \textit{2.94 $\pm$ 0.10} & $6.36 \pm 0.47$ & \textbf{6.13 $\pm$ 0.09} & $67.71 \pm 5.09$ & $422.11$ & $189.0$ & $1113.7$\\
								      & $86400s$ & \textit{4.98 $\pm$ 0.04} & $53.74 \pm 1.18$ & \textbf{43.81 $\pm$ 0.40} & $115.91 \pm 1.68$ & $421.34$ & $4287.7$ & $18045.1$  \\ 
								      & $259200s$ & \textit{8.27 $\pm$ 0.03} & $340.90 \pm 1.59$ & $278.19 \pm 5.52$ & \textbf{198.51 $\pm$ 5.89} & $424.83$ & $13804.3$ & $55494.2$ \\
	\midrule
	\multirow{3}{*}{RE} & $3600s$ & \textit{16.81 $\pm$ 0.03} & $71.36 \pm 2.26$ & \textbf{69.73 $\pm$ 1.99} & $570.66 \pm 50.26$ & $18656.79$ & $1708.7$ & $6406.0$ \\
								       & $86400s$ & \textit{29.75 $\pm$ 0.14} & $111.83 \pm 0.69$ & \textbf{103.45 $\pm$ 4.35} & $943.05 \pm 67.62$ & $18528.12$ & $7479.5$ & $23085.0$  \\ 
								       & $259200s$ &\textit{50.66 $\pm$ 0.39} & $192.88 \pm 2.82$ & \textbf{164.89 $\pm$ 2.08} & $1121.81 \pm 79.14$ & $18949.97$ & $11165.8$ & $29680.1$ \\
	\midrule
	\multirow{3}{*}{EC} & $1 \times 10^5 \mu s$ & \textit{149.75 $\pm$ 5.45} & \textbf{372.04 $\pm$ 0.18} & $425.55 \pm 16.79$ & \xmark & \xmark & \xmark & \xmark \\
									   & $2 \times 10^5 \mu s$ & \textit{180.04 $\pm$ 5.81} & \textbf{441.08 $\pm$ 19.44} & $497.36 \pm 8.64$ & \xmark & \xmark & \xmark & \xmark \\
									   & $3 \times 10^5 \mu s$ & \textit{207.54 $\pm$ 1.11} & \textbf{473.33 $\pm$ 0.73} & $580.25 \pm 0.86$ & \xmark & \xmark & \xmark & \xmark \\
	\bottomrule
	\end{tabular}
}
\end{table}

	\section{Predictors correlation}
	\label{appendix::predictors_correlation}
	
		\begin{figure}
		\includegraphics[width=\textwidth]{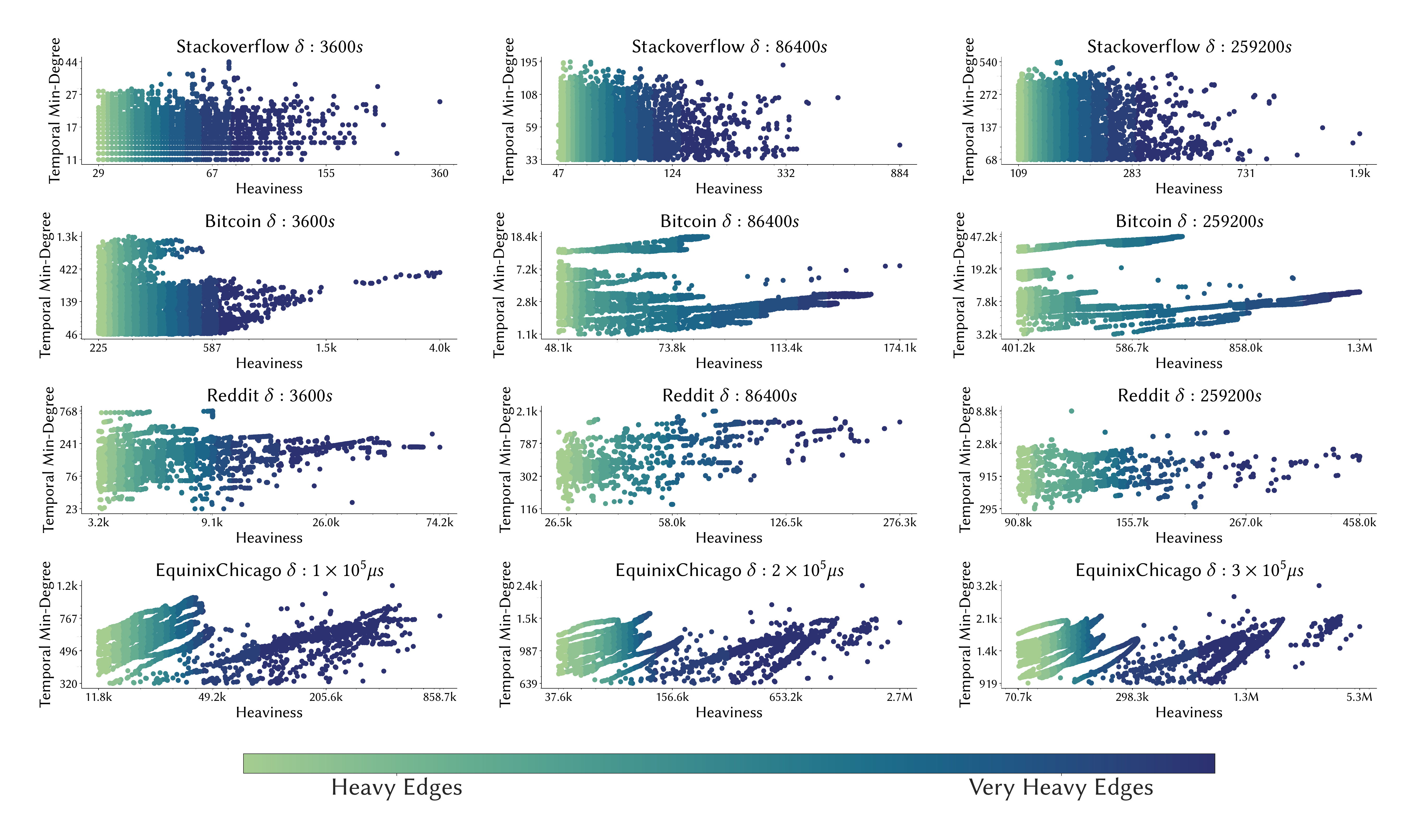}
		\caption{Correlation between perfect predictor and temporal min-degree predictor on all datasets and for all values of $\delta$. Each point represents a temporal edge with a given heaviness (x-axis) and a corresponding minimum temporal degree (y-axis). Both the y and the x axes are in log-scale. We selected the top-$1 \times 10^4$ heaviest edges for visualization purposes.}
		\label{fig:predictors_correlation_offline}
	\end{figure}
	
	\begin{figure}
		\includegraphics[width=\linewidth]{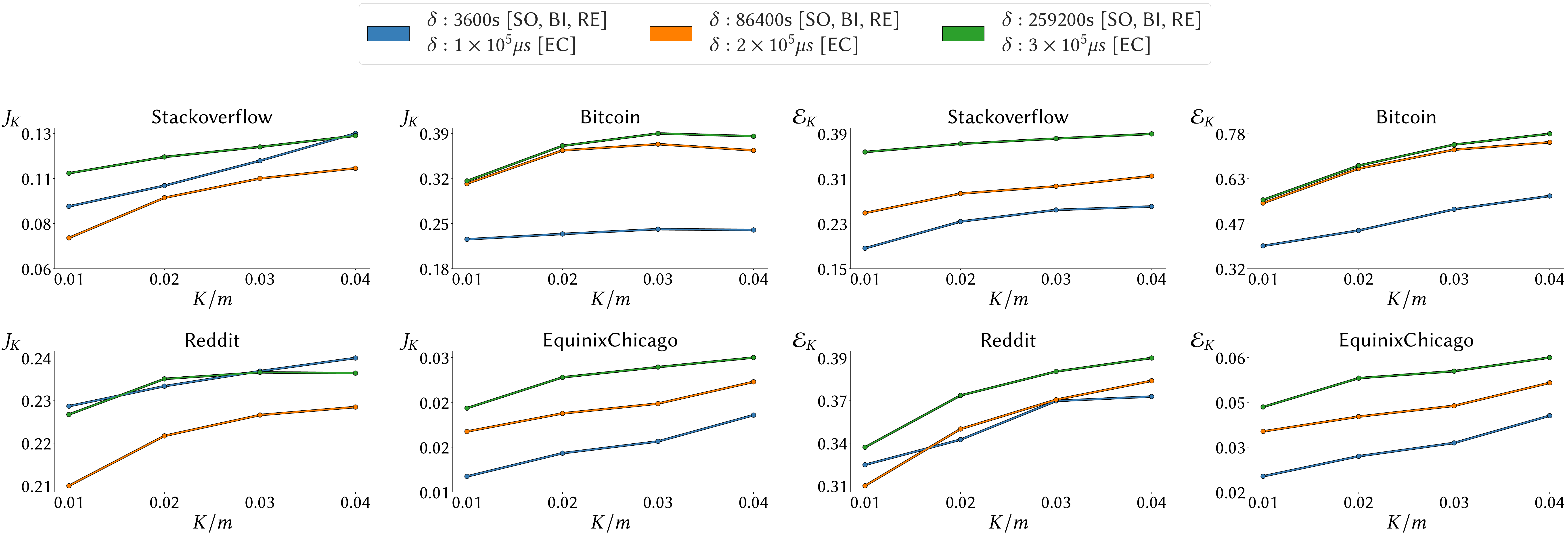}
		\caption{Left: Jaccard similarity between the set \topKPerfect and the set \topKTemporalDeg{K'} as a function of $K/m$. Right: \VMetric metric between the set \topKPerfect and the set \topKTemporalDeg{K'} as a function of $K/m$.}
		\label{fig:predictors_correlation_online}
	\end{figure}
	
	In this section, we study the correlation between a perfect predictor and the temporal min-degree predictor in both of the scenarios considered, i.e., predictor learned on the whole data and predictor learned on past data for online estimation. We will denote with $\topKPerfect \doteq \langle e^\totEdgeW_1,\dots, e^\totEdgeW_{K} \rangle$ the top-$K$ edges in $E$ according to the weight \totEdgeWeight{\cdot} and with $\topKTemporalDeg{K'} \doteq \langle e^\totEdgeW_1,\dots, e^\totEdgeW_{K'} \rangle$  the top-$K'$ edges in $E$ according to the temporal min-degree weight \edgeTemporalDeg{\cdot}.
	
	\spara{Predictor learned on the whole data}. In this scenario, we retrieve the edges \topKPerfect and the edges \topKTemporalDeg{K'} considering the whole input graph; we also fix $K = K'$. \Cref{fig:predictors_correlation_offline} shows the relation between the temporal min-degree weight \edgeTemporalDeg{e} of an edge $e$ and its weight \totEdgeWeight{e}. For each dataset and for each time-duration $\delta$, we selected the top-$1 \times 10^4$ heaviest edges that are in $\topKPerfect \cap \topKTemporalDeg{K}$. 
	On the BI and RE datasets, the correlation between the two rankings is clearly visible, especially for larger values of $\delta$ over dataset BI. 
	For the SO dataset, there is not a clear trend, and many light edges may seem to be classified as heavy by the temporal min-degree predictor (e.g., see SO in \Cref{fig:predictors_correlation_offline} for $\delta=259\,200s$). 
	On the EC dataset, we can see that there is a subset of heavy edges that have a high temporal min-degree weight (especially for small $\delta$), which are likely to be ranked as heavy by the temporal min-degree predictor. 
	However, a non-negligible amount of  heavy edges that have a low temporal min-degree weight is not classified correctly, which likely reduces the accuracy of our temporal min-degree predictor. 
	This is particularly evident for high values of $\delta$. 
	Nevertheless, our implementation of the temporal min-degree predictor is quite effective in the scenario we considered even for the EC dataset, as discussed in \Cref{sec:soa_comp}.
	
	\spara{Predictor learned on historical data}. We now focus on the second scenario considered, where the set \topKPerfect is derived only from the test stream $\tau^{ts}$ and the set \topKTemporalDeg{K'} comprises the edges that are classified as heavy according to their temporal min-degree weight using a threshold \predThreshold learned on past data (i.e., the training stream $\tau^{tr}$).
	In this setting, we have $K \neq K'$ in general. It is important to state that \predThreshold is learned on the top-$K$ edges according to the temporal min-degree weight, i.e., the parameter $K$ has an impact on both the set \topKTemporalDeg{K'} and \topKPerfect. In particular, the value $K'$ is in direct relationship with $K$ since an increase (decrease) in $K$ produces an increase (decrease) in $K'$. We used two correlation metrics in our analysis:
	\begin{enumerate}
		\item[(1)] the Jaccard similarity  $\jaccardSim = \frac{|\topKTemporalDeg{K'} \cap \topKPerfect|}{|\topKTemporalDeg{K'} \cup \topKPerfect|}$; 	
		\item[(2)] the \VMetric metric, defined as $\VMetric = \frac{|\topKTemporalDeg{K'} \cap \topKPerfect|}{|\topKTemporalDeg{K'}|}$.
	\end{enumerate}
	
	The Jaccard similarity \jaccardSim measures the percentage of the top-$K$ heavy edges according to $\totEdgeW{(\cdot)}$ that are captured by the temporal min-degree predictor. The \VMetric metric measures instead the percentage of heavy edges (according to the weight $\totEdgeW{(\cdot)}$) that are included in \topKTemporalDeg{K'} by the temporal min-degree predictor. For each dataset and time-duration $\delta$, we computed the value of \jaccardSim and \VMetric for different values of $K$. The results reported in \Cref{fig:predictors_correlation_online} show that the temporal min-degree predictor is particularly effective for the BI and RE datasets, especially for the largest $\delta$. In SO and EC datasets, the correlation values are smaller, with the ones of EC being one order of magnitude smaller than all other datasets, in agreement with the results obtained in the first scenario.

\end{document}